\documentclass[article]{jss}

\usepackage{orcidlink,thumbpdf,lmodern}
\usepackage{multicol, tikz}
\usepackage{framed,amsmath,amssymb}
\usepackage{algorithm}
\usepackage{algpseudocode}

\defcitealias{Miller}{Miller et al. (2005)}
\defcitealias{bioconductor}{Gentleman et al. (2004)}

\author{Adam Chojecki\\Warsaw University of Technology \And
	Pawe{\l} Morgen\\Warsaw University of Technology \AND
	Bartosz Ko{\l}odziejek~\orcidlink{0000-0002-5220-9012}\\Warsaw University of Technology}
\Plainauthor{Adam Chojecki, Pawe{\l} Morgen, Bartosz Ko{\l}odziejek}

\title{Learning Permutation Symmetry\\of a Gaussian Vector with \pkg{gips} in \proglang{R}}
\Plaintitle{Learning Permutation Symmetries with gips in R}
\Shorttitle{\pkg{gips}: Learning Permutation Symmetries}

\Abstract{
	The study of hidden structures in data presents challenges in modern statistics and machine learning. We introduce the \pkg{gips} package in \proglang{R}, which identifies permutation subgroup symmetries in Gaussian vectors. \pkg{gips} serves two main purposes: exploratory analysis in discovering hidden permutation symmetries and estimating the covariance matrix under permutation symmetry. It is competitive to canonical methods in dimensionality reduction while providing a new interpretation of the results. \pkg{gips} implements a novel Bayesian model selection procedure within Gaussian vectors invariant under the permutation subgroup introduced in \cite{GIKM}, The Annals of Statistics, 50 (3) (2022).
}

\Plainkeywords{Bayesian model selection, permutation symmetry, high-dimensional statistics, dimensionality reduction, parameter sharing, R}

\Address{
	Adam Chojecki, Pawe{\l} Morgen, Bartosz Ko{\l}odziejek\\
	Warsaw University of Technology\\
	Faculty of Mathematics and Information Science\\
	Koszykowa 75\\ 
	00-662 Warsaw, Poland\\
	E-mail: \email{adam.prze.choj@gmail.com}, \email{seriousmorgen@protonmail.com}, \\
	\email{bartosz.kolodziejek@pw.edu.pl}
}

\begin{document}
	
	\section[Introduction]{Introduction} \label{sec:intro}
	
	The study of hidden structures in the data is one of the biggest challenges in modern mathematical statistics and machine learning \cite{ElementsOfStatLearn}. 
	Extracting meaningful information from high-dimensional datasets, where the number of variables $p$ exceeds the number of observations $n$, poses a significant hurdle due to the curse of dimensionality.
	
	One solution to the problem of an insufficient number of observations relative to the number of variables is to restrict to models with lower dimensionality. Graphical models have been introduced for this purpose \cite{L96}, where a conditional independence structure (graph Markovian structure) is imposed on the distribution of a random vector. Such structures are conveniently described by graphs and allow for a reduction in the dimensionality of the problem. However, if the graph is not sparse enough, then such a procedure does not allow for a reliable estimation of the covariance matrix. We note that the study of the covariance matrix is the basic way to describe the dependency structure of a random vector and provides a convenient way to quantify the dependencies between variables.
	
	If the data is insufficient and some inference must be performed, one has to propose additional assumptions or restrictions. In such a situation, colored graphical models could be considered, where, in addition to conditional independence, certain equality conditions on the covariance matrix are imposed. 
	Incorporating such equality conditions in colored graphical models is an example of parameter sharing. This concept allows for a reduction of dimensionality and can effectively incorporate domain knowledge into the model architecture. A notable example of parameter sharing, which possesses these advantages, is the convolution technique \cite{ElementsOfStatLearn} used in image and video processing, enabling efficient feature extraction and pattern recognition.
	
	A rich family of such symmetry conditions can be expressed using the language of permutations. This idea was introduced in \cite{AnderssonInvariant, AndersonPermSymmetry} and \cite{HL08}.
	In the latter paper, three types of such models (RCOP among them) were introduced to describe situations where some entries of concentration or partial correlation matrices are approximately equal. These equalities can be represented by a colored graph. The RCOP model, apart from the graph Markovian structure, permits additional invariance of the distribution with respect to some permutation subgroup. We say that the distribution of a $p$-dimensional random vector $Z$ is invariant under permutation subgroup $\Gamma$ on $V=\{1,\ldots,p\}$ if $Z=(Z_i)_{i\in V}$ has the same distribution as $(Z_{\sigma(i)})_{i\in V}$ for any permutation $\sigma\in\Gamma$, \cite{AnderssonInvariant}. This property is called the permutation symmetry of the distribution of $Z$ and imposes significant symmetry conditions on the model.
	
	The case when the conditional dependency graph is unknown or known to be the complete graph was studied in \cite{GIKM}. In that paper, the authors introduced a Bayesian model selection procedure for the case when $Z$ is a Gaussian vector. In other words, by assuming a prior distribution on the parameters, they derived the posterior probability of a specific model. This allows one to find the permutation group under which (most likely) the data is invariant. Not only does this result in dimensionality reduction but also provides a simple and natural interpretability of the results. For example, if the distribution of $Z$ is invariant under swapping its $i$th and $j$th entries, then one can say that both $Z_i$ and $Z_j$ play a symmetrical role in the model.
	
	The concept of group invariance finds application in various domains and often leads to improved estimation properties. If the group under which the model is invariant is known, precise convergence rates for the regularized covariance matrix were derived in \cite{Shah12}, demonstrating significant statistical advantages in terms of sample complexity. Another noteworthy paper, \cite{Solo16}, explores group symmetries for estimating complex covariance matrices in non-Gaussian models, which are invariant under a known permutation subgroup. However, neither of these articles provides guidance on identifying the permutation subgroup when it is unknown, which is typically the case in practical applications. 
	
	Identifying the permutation subgroup symmetry can be interpreted as an automated way of extracting expert knowledge from the data. Discovering the underlying symmetries allows for a deeper understanding of the relationships and dependencies between variables, offering insights that may not be apparent through traditional analysis alone. The automated approach reduces the reliance on manual exploration and expert intervention.
	
	In the present paper, we introduce an \proglang{R} package called \pkg{gips} (acronym derived from 'Gaussian model invariant by permutation symmetry'), \cite{gips}, which implements the model selection procedure described in \cite{GIKM}. The \pkg{gips} package, presented in this paper, serves two purposes:
	\begin{enumerate}
		\item Discovering hidden permutation symmetries among variables (exploratory analysis).
		\item Estimating covariance matrix under the assumption of known permutation symmetry.
	\end{enumerate}
	Both points are limited to the Gaussian setting. To the best of our knowledge, there are currently no other software packages available (in \proglang{R} or any other programming language) that address the topic of finding permutation symmetry. Our approach focuses on zero-mean Gaussian vectors, although the method can be applied to centered data and, if the sample size $n$ is reasonably large, to standardized data as well, see Section \ref{sec:standardizing}.
	
	Let $Z$ be a Gaussian vector with a known mean. If we assume full symmetry of the model, see Section \ref{sec:permSymmetry}, meaning that the distribution of $Z$ is invariant under any permutation, then the maximum likelihood estimator (MLE) of the covariance matrix requires only a single sample ($n_0=1$) to exist. Somewhat surprisingly, the same phenomenon applies when the normal sample is invariant under a cyclic subgroup generated by a cycle of length $p$. While it is natural to consider permutation symmetries alongside conditional independence structures, we follow \cite{GIKM} and assume no conditional independencies among the variables. Such an approach already enables a substantial reduction in dimensionality, accompanied by a readily interpretable outcome. The development of the method to incorporate non-trivial graph Markovian structures is a topic for future research, and we will consider expanding the package if a new theory emerges. The first step towards generalizing the theory to homogeneous graphs has already been taken in \cite{GKI22}. Additionally, a simple heuristic can be employed to identify non-trivial Markovian structures using our model - see \cite[Section 1.2]{GIKM}, \cite[Section 4.1]{GIKM_SM}, and Section \ref{sec:brease_cancer} in this paper.

	Although there are no other software packages available for finding permutation symmetries in data, we have made the decision to compare the results of our model with canonical methods commonly used to tackle high-dimensional problems, namely Ridge and Graphical LASSO (GLASSO) estimation and model selection (implemented, for instance, in \proglang{R} packages: \pkg{huge} \cite{huge} and \pkg{rags2ridges} \cite{rags2ridges, rags2ridges_paper}). These methods correspond to estimation with constraints or, conversely, to Bayesian estimation with Gaussian or Laplace priors, respectively \cite[Sec. 6.2.2]{ISL}. We demonstrate that \pkg{gips} is competitive with these widely used approaches in terms of dimensionality reduction properties, and moreover, it offers interpretability of the results in terms to permutation symmetries.
	
	Furthermore, it is worth noting that due to the discrete nature of the problem, we believe that finding permutation symmetry cannot be adequately addressed by penalized likelihood methods, which are generally much faster than Bayesian methods. Although other methods (which do not have available implementations to our best knowledge) allow for model selection within colored graphical Gaussian models, none of them are applicable to permutation invariant models (RCOP models). Compared to other models (such as RCON, RCOR in \cite{HL08}, which correspond to different type of restrictions/symmetries), RCOP models possess a more elegant algebraic description and offer a natural interpretation \cite{Ge11,GM15, MassamBayesian,Hel20,QXNX21,RRL21}.
	
	The ``Replication code'' is available at \url{https://github.com/PrzeChoj/gips_replication_code}. 
	
	\subsection{Overview of the paper}
	
	The paper is organized as follows. The Introduction consists of four subsections. In the next subsection, we present two low-dimensional toy examples that illustrate the use of \pkg{gips}.
	In the subsequent subsection, we discuss the potential for successfully exploiting group symmetry in many natural real-life problems. 
	In the final subsection of the Introduction, we argue that it is both necessary and sufficient to focus on cyclic symmetries, which are more tractable.
	
	Section \ref{sec:models} provides the necessary methodological background on permutation symmetries and defines the Bayesian model proposed in \cite{GIKM}, specialized to cyclic subgroups. We also introduce an Markov chain Monte Carlo (MCMC) algorithm that allows the estimation of the maximum a posteriori (MAP) within our Bayesian model, and we discuss the issue of centering and standardizing input data.
	
	Section \ref{sec:illustrations} is dedicated to numerical simulations. We present a high-dimensional example using breast cancer data from \citetalias{Miller}. Additionally, we use a heuristic approach from \cite{GIKM} for identifying the graphical model invariant under permutation symmetries (RCOP model from \cite{HL08}) and apply this procedure to the real-life example. In the subsequent subsections, we examine the impact of hyperparameters on model selection and compare \pkg{gips} with competing packages that facilitate dimensionality reduction.
	
	Finally, in Section \ref{sec:summary} we draw some conclusions.
	
	An example to Section \ref{sec:vs} is presented in Appendix \ref{app:general}.
	Mathematical details behind the Bayesian model are relegated to the Appendix \ref{app:technical}.

	\subsection{Toy examples} \label{sec:toyexps}

	We illustrate the concept of permutation symmetry using the \pkg{gips} package in two simple use cases. These examples demonstrate how permutational symmetry can enhance the data mining process. A similar procedure was successfully applied to the Frets' heads dataset \cite[Section 4.2]{GIKM} and the mathematical marks dataset \cite[Section 4]{GKI22}.
	
	In the first example, we use \code{aspirin} dataset from the \pkg{HSAUR2} package. By examining the covariance matrix, we manually choose a reasonable permutation symmetry. Additionally, we employ the \pkg{gips} package to demonstrate that our algorithm generates reasonable estimates.
	
	For the second example, we utilize the \code{oddbooks} dataset from the \pkg{DAAG} package. We showcase how one can incorporate expert field knowledge in the analysis. We use the \pkg{gips} to find the permutation symmetry and interpret the result.
	
	A standard personal computer (PC) can execute the entire code in this section within $10$ seconds.
	
	\subsubsection{Aspirin dataset}
	
	This dataset consists of information about a meta-analysis of the efficacy of Aspirin (versus placebo) in preventing death after a myocardial infarct.
	
	We renumber the columns for better readability:
	\begin{CodeChunk}
		\begin{CodeInput}
R> data("aspirin", package = "HSAUR2")
R> Z <- aspirin
R> Z[, c(2, 3)] <- Z[, c(3, 2)]
R> names(Z) <- names(Z)[c(1, 3, 2, 4)]
R> head(Z, 4)
		\end{CodeInput}
		\begin{CodeOutput}
   dp  da  tp  ta
1  67  49 624 615
2  64  44 771 758
3 126 102 850 832
4  38  32 309 317
		\end{CodeOutput}
	\end{CodeChunk}
	
	Each of the $n=7$ rows in \code{Z} corresponds to a different study, and the $p=4$ columns represent the following: \code{dp}: number of deaths after placebo, \code{da}: number of deaths after Aspirin, \code{tp}: total number subjects treated with placebo, \code{ta}: total number of subjects treated with Aspirin.
	
	Initially, we calculate the empirical covariance matrix \code{S}:
	
	\begin{CodeChunk}
		\begin{CodeInput}
R> n <- nrow(Z)
R> p <- ncol(Z)
R> S <- cov(Z)
		\end{CodeInput}
	\end{CodeChunk}
	
\begin{figure}[h]
		\centering
		\includegraphics[width=0.7\textwidth]{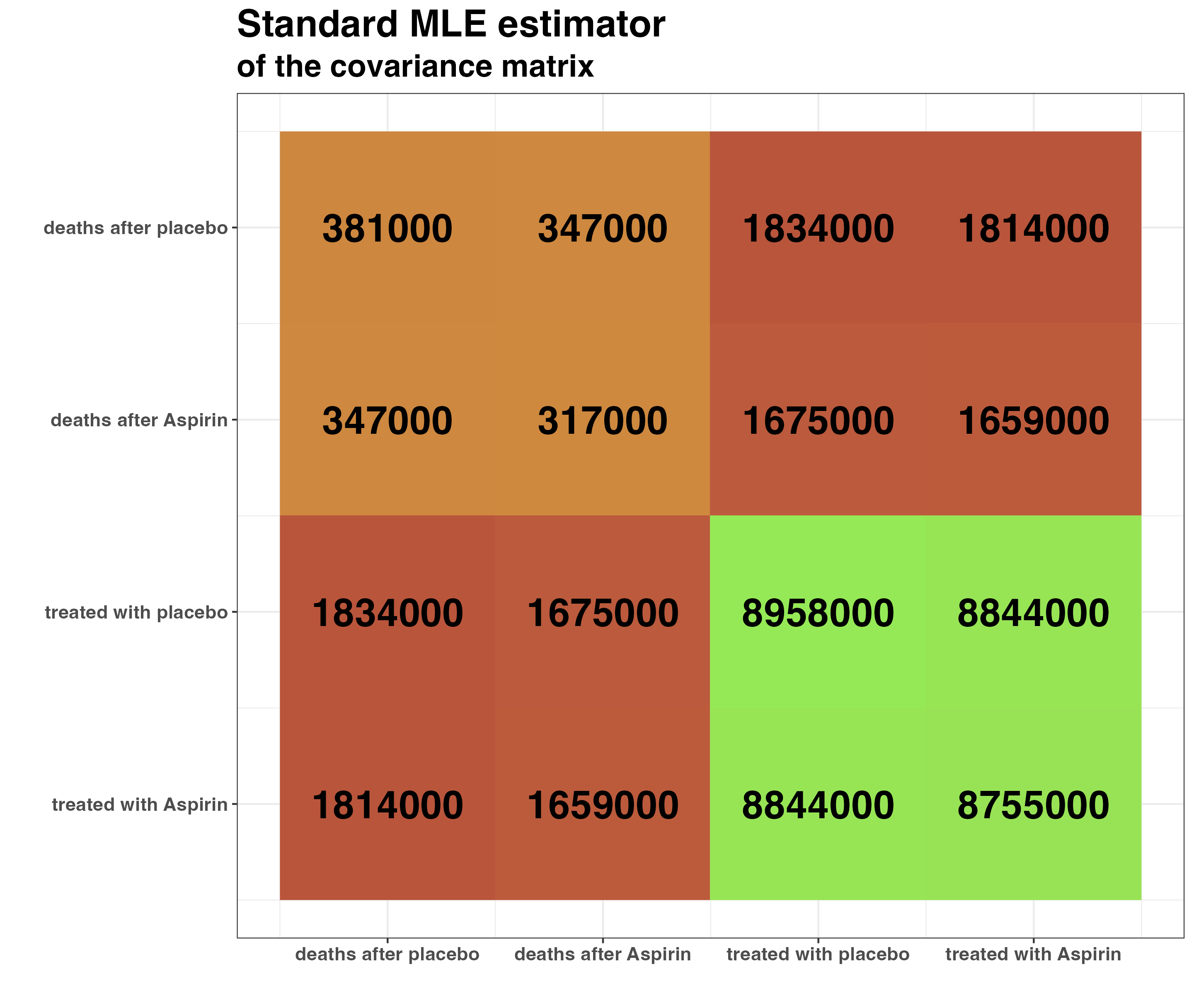}
		\caption{\small Standard MLE of the covariance matrix of the \code{aspirin} dataset. We can see that some entries of \code{S} have similar colors, which suggests a lower dimensional model with equality constraints.}
		\label{fig:aspirin_id}
	\end{figure}

	Note that since $n=7$ is greater than $p=4$, \code{S} is the standard MLE of $\Sigma$ in the (unrestricted) Gaussian model. The heatmap of the \code{S} matrix is shown in Figure~\ref{fig:aspirin_id}.
	
	We observe significant similarities between the empirical covariances of variables \code{tp} (column $3$) and \code{ta} (column $4$). They exhibit comparable variances (\code{S[3,3]} $\approx$ \code{S[4,4]}), and their covariances with the other variables also show resemblance (\code{S[1,3]} $\approx$ \code{S[1,4]} and \code{S[2,3]} $\approx$ \code{S[2,4]}). 
	
	By definition, the distribution of a random vector $Z=(Z_1,Z_2,Z_3,,Z_4)^\top$ is invariant under the permutation $(3,4)$ if the distributions of $(Z_1,Z_2,Z_3,Z_4)^\top$ and $(Z_1, Z_2, Z_4, Z_3)^\top$ coincide. When $Z$ follows a centered Gaussian distribution, this property can be expressed purely in terms of its covariance matrix, leading to the following conditions: $\VAR[Z_3] = \VAR[Z_4]$, $\COV[Z_1, Z_3] = \COV[Z_1, Z_4]$, and $\COV[Z_2, Z_3] = \COV[Z_2, Z_4]$.
	We observe that the structure of \code{S} closely corresponds to that of the covariance matrix of a random vector invariant under the permutation $(3,4)$. By observing that \code{S[1,1]} $\approx$ \code{S[2,2]}, \code{S[1,3]} $\approx$ \code{S[2,3]}, and \code{S[1,4]} $\approx$ \code{S[2,4]}, we can also argue that the data is invariant under the permutation $(1,2)$ or even $(1,2)(3,4)$.
	
	We want to emphasize that such manual exploration becomes infeasible for larger values of $p$ due to the massive number and complexity of possible relationships.
	A priori, it is unclear which scenario is preferable (one can compare Bayesian information criterion (BIC), but the MLE does not always exist). 
	The \pkg{gips} package uses the Bayesian paradigm (described in detail in Section \ref{sec:BM}) to precisely quantify posterior probabilities of considered permutation groups. 
	The workflow in \pkg{gips} is as follows: first, use the \code{gips()} function to define an object of the class \code{`gips`} that contains all the necessary information for the model. Next, use the \code{find_MAP()} function with an optimizer of your choice to find the permutation that provides the maximum a posteriori estimate. Finally, we use the \code{project_matrix()} function to obtain the MLE of the covariance matrix in the invariant model, which will serve as a more stable covariance estimator. The process can be summarized as follows:
	\begin{CodeChunk}
		\begin{CodeInput}
R> g <- gips(S, n)
R> g_MAP <- find_MAP(g, optimizer = "BF",
+    save_all_perms = TRUE, return_probabilities = TRUE
+  )
R> g_MAP
		\end{CodeInput}
		\begin{CodeOutput}
The permutation (1,2)(3,4):
 - was found after 17 posteriori calculations;
 - is 3.374 times more likely than the () permutation.
		\end{CodeOutput}
	\end{CodeChunk}
	According to the output of \code{find_MAP()}, the permutation $(1,2)(3,4)$ best reflects the symmetries of the models and is over $3$ times more probable (under our Bayesian setting) than the identity permutation \code{()}, which corresponds to no symmetry.
	The invariance with respect to the permutation $(3,4)$ arises from the fact that the samples of patients treated with aspirin and placebo had similar sizes. On the other hand, the invariance with respect to the permutation $(1,2)$ signifies the lack of aspirin treatment effect. The permutation $(1,2)(3,4)$ corresponds to both of these effects. We emphasize that this study is an exploratory analysis rather than a statistical test.
	
	We can easily calculate probabilities of all symmetries using a built-in function:
	\begin{CodeChunk}
		\begin{CodeInput}
R> get_probabilities_from_gips(g_MAP)
		\end{CodeInput}
		\begin{CodeOutput}
  (1,2)(3,4)        (3,4)        (1,2)           ()        (1,4)        (1,3) 
5.107108e-01 1.695605e-01 1.663982e-01 1.513854e-01 4.341644e-04 4.047690e-04 
       (2,4)        (2,3)    (1,3,2,4)   (1,3)(2,4)   (1,4)(2,3)      (1,3,4) 
3.797581e-04 3.607292e-04 1.240381e-04 7.410652e-05 7.406484e-05 2.197791e-05 
     (1,2,4)      (1,2,3)      (2,3,4)    (1,2,4,3)    (1,2,3,4) 
2.026609e-05 1.813565e-05 1.782315e-05 7.676231e-06 7.528912e-06 
		\end{CodeOutput}
	\end{CodeChunk}
	or compare two permutations of interest:
	\begin{CodeChunk}
		\begin{CodeInput}
R> compare_posteriories_of_perms(g_MAP, "(34)")
		\end{CodeInput}
		\begin{CodeOutput}
The permutation (1,2)(3,4) is 3.012 times more likely
than the (3,4) permutation.
		\end{CodeOutput}
		\begin{CodeInput}
R> compare_posteriories_of_perms(g_MAP, "(12)")
		\end{CodeInput}
		\begin{CodeOutput}
The permutation (1,2)(3,4) is 3.069 times more likely
than the (1,2) permutation.
		\end{CodeOutput}
		\begin{CodeInput}
R> compare_posteriories_of_perms(g_MAP, "()")
		\end{CodeInput}
		\begin{CodeOutput}
The permutation (1,2)(3,4) is 3.374 times more likely
than the () permutation.
		\end{CodeOutput}
	\end{CodeChunk}
	
	Note that for $p=4$, there are $p!=24$ different permutations, but only $17$ distinct symmetries are reported above. This is because some permutations correspond to the same symmetry. More precisely, it is the group generated by a permutation $\sigma$ and not $\sigma$ itself that identifies the symmetry. For example $\sigma_1 = (1,2,3)$ and $\sigma_2 = (1,3,2)$ generate the same group.
	
	We also note that given the small number of variables ($p=4$), the space of possible permutation symmetries is also small. Consequently, we were able to compute the exact posterior probabilities of our Bayesian model for every single permutation symmetry. The number of permutation symmetries grows superexponentially with $p$, e.g., for $p=10$ its cardinality is approximately $1$ million (see OEIS\footnote{The On-Line Encyclopedia of Integer Sequences, \url{https://oeis.org/}.} sequence A051625). Thus, for larger $p$ we recommend using the implemented Metropolis-Hastings algorithm to approximate these probabilities, see Section \ref{sec:search}.
	
\begin{figure}[h]
		\centering
		\includegraphics[width=0.7\textwidth]{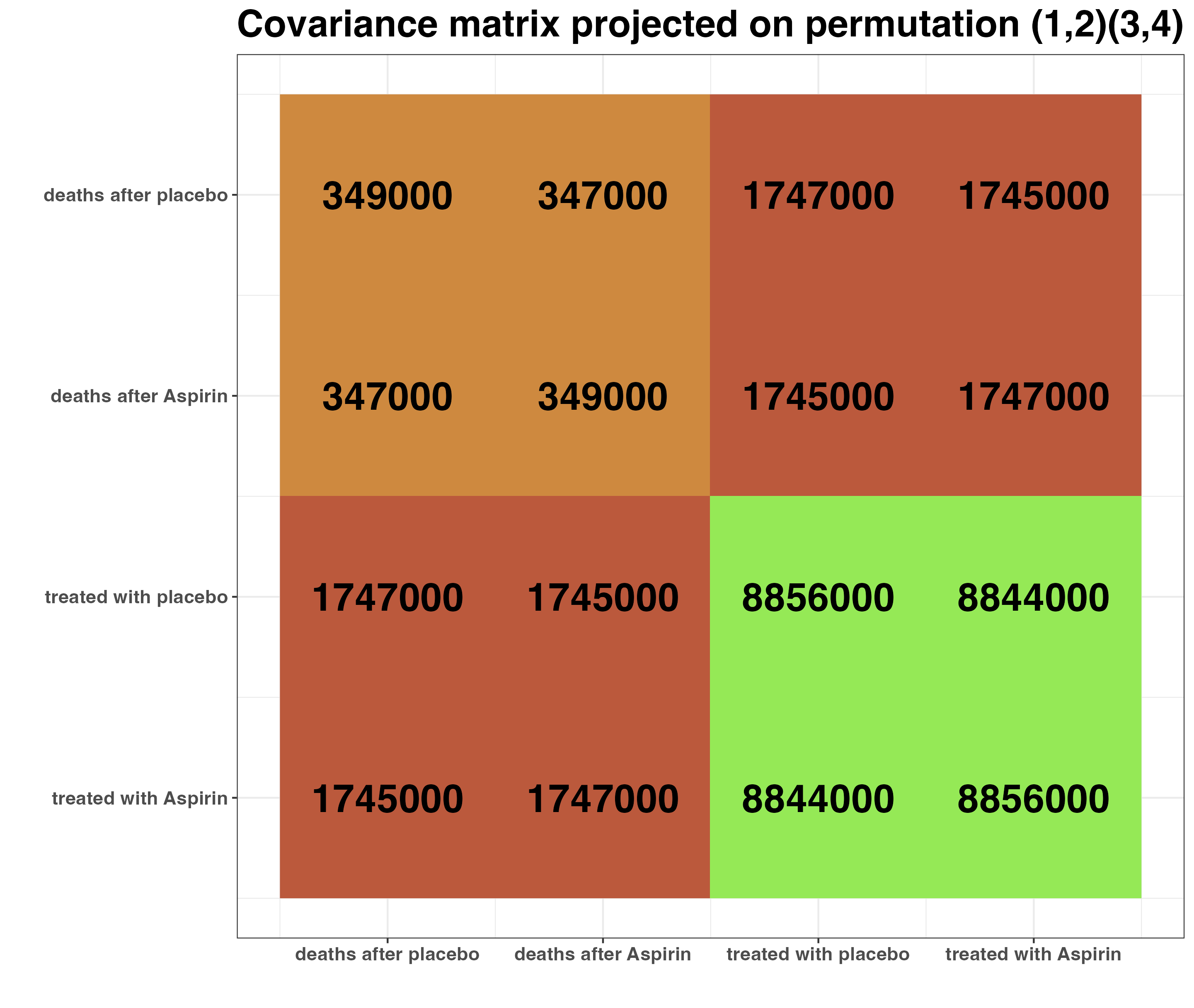}
		\caption{\small The MLE of the covariance matrix of the \code{oddbooks} dataset in the model invariant under the permutation $(1,2)(3,4)$. We can see that the entries corresponding to similar colors in Figure~\ref{fig:aspirin_id} are exactly equal.}
		\label{fig:aspirin_map}
	\end{figure}

	Assuming that the data actually come from a distribution invariant under the permutation $(1,2)(3,4)$, we can provide a new estimate for the covariance matrix. Formally, we project the matrix \code{S} onto the space of positive definite matrices that are invariant under the permutation $(1,2)(3,4)$ (for further details, refer to Section \ref{sec:MLE}). In practice, we enforce the desired equalities by averaging.
	\begin{CodeChunk}
		\begin{CodeInput}
R> S_projected <- project_matrix(S, g_MAP)
		\end{CodeInput}
	\end{CodeChunk}
	One can easily plot the found covariance estimator with a line:
	\begin{CodeChunk}
		\begin{CodeInput}
R> plot(g_MAP, type = "heatmap")
		\end{CodeInput}
	\end{CodeChunk}
	It is shown in Figure~\ref{fig:aspirin_map} (we made cosmetic modifications to this plot; the exact code is provided in the attached ``Replication code'').
	
	The \code{S_projected} matrix can now be interpreted as a more stable covariance matrix estimator, see e.g., \cite{Shah12, Solo16}.

	\subsubsection{Books dataset}
	
	This dataset consists of information about thickness (mm), height (cm), width (cm), and weight~(g) of $12$ books.
	
	\begin{CodeChunk}
		\begin{CodeInput}
R> data("oddbooks", package = "DAAG")
R> head(oddbooks, 4)
		\end{CodeInput}
		\begin{CodeOutput}
   thick height breadth weight
1     14   30.5    23.0   1075
2     15   29.1    20.5    940
3     18   27.5    18.5    625
4     23   23.2    15.2    400
		\end{CodeOutput}
	\end{CodeChunk}
	
	We will only consider relationships between the thickness, height, and width.
	\begin{CodeChunk}
		\begin{CodeInput}
R> Z <- oddbooks[, c(1, 2, 3)]
		\end{CodeInput}
	\end{CodeChunk}
	
	One can suspect that books from this dataset were printed with a $\sqrt{2}$ aspect ratio, as in the popular A-series paper size. Therefore, we can utilize this domain knowledge in the analysis and unify the data for height and width:
	\begin{CodeChunk}
		\begin{CodeInput}
R> Z$height <- Z$height / sqrt(2)
		\end{CodeInput}
	\end{CodeChunk}
	
\begin{figure}[h]
		\centering
		\includegraphics[width=0.7\textwidth]{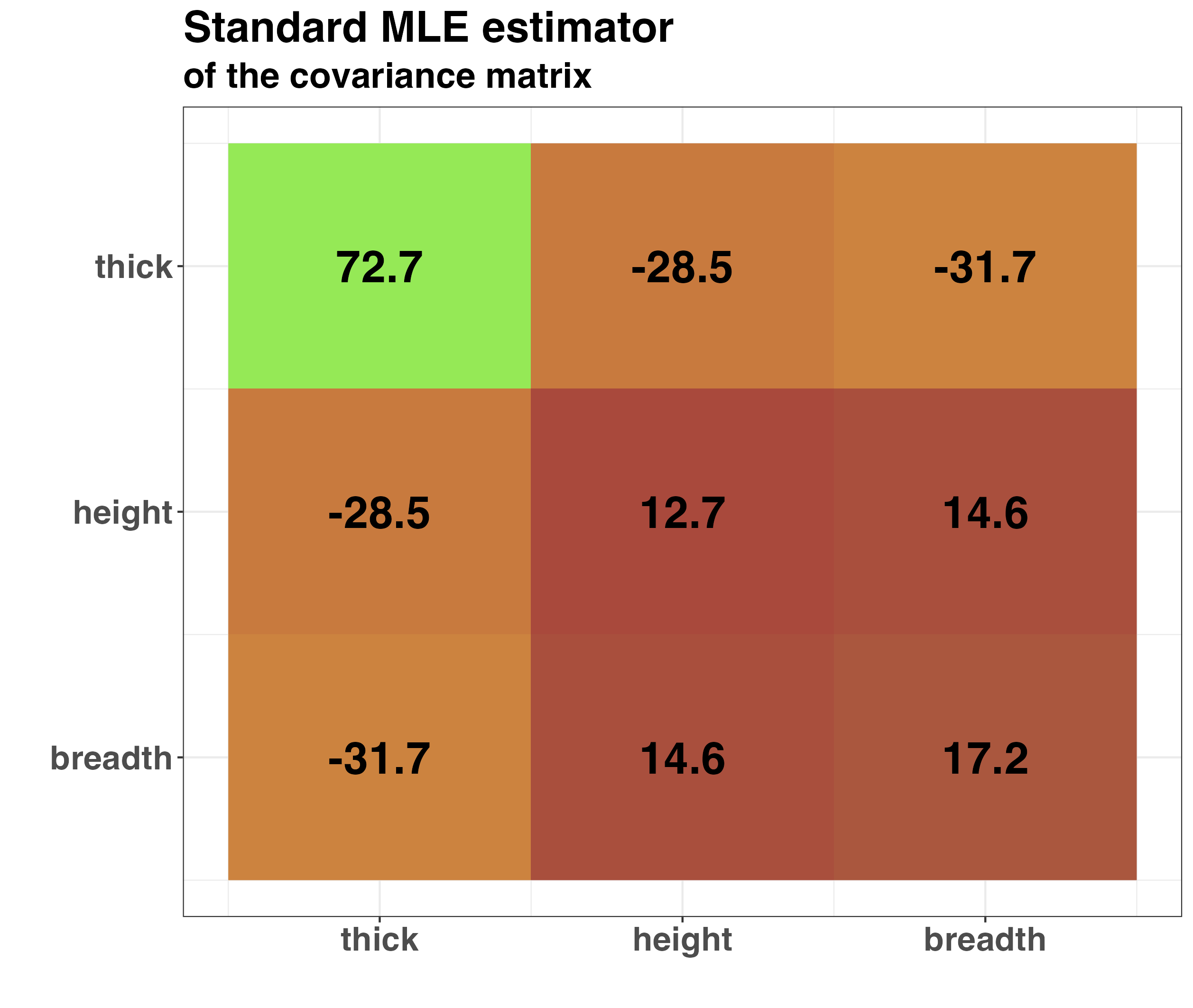}
		\caption{\small Standard MLE of the covariance matrix of the \code{oddbooks} dataset. We can see that some
			entries of \code{S} have similar colors, which suggests a lower dimensional model with equality constraints.}
		\label{fig:books_id}
	\end{figure}

	Let us see the standard MLE of the covariance matrix:
	\begin{CodeChunk}
		\begin{CodeInput}
R> S <- cov(Z)
		\end{CodeInput}
	\end{CodeChunk}
	We can plot this covariance matrix to see if we would notice any connection between variables. Figure~\ref{fig:books_id} was obtained with the code below (we made cosmetic modifications to this plot; the exact code is provided in the attached ``Replication code''):
	
	\begin{CodeChunk}
		\begin{CodeInput}
R> g <- gips(S, number_of_observations)
R> plot(g, type = "heatmap")
		\end{CodeInput}
	\end{CodeChunk}
	
	We can see that some
	entries of \code{S} have similar colors, which suggests a lower dimensional model with equality constraints. In particular, the covariance between \code{thick} and \code{height} is very similar to the covariance between \code{thick} and \code{breadth}, and the variance of \code{height} is similar to the variance of \code{breadth}. Those are not surprising, given the data interpretation (after the \code{height} rescaling that we did).
	
	Let us examine the posterior probabilities returned by \pkg{gips}:
	\begin{CodeChunk}
		\begin{CodeInput}
R> g_MAP <- find_MAP(g, optimizer = "BF",
+    return_probabilities = TRUE, save_all_perms = TRUE
+  )
R> get_probabilities_from_gips(g_MAP)
		\end{CodeInput}
		\begin{CodeOutput}
       (2,3)           ()        (1,3)      (1,2,3)        (1,2) 
5.660781e-01 4.339087e-01 6.728772e-06 4.683290e-06 1.862353e-06
		\end{CodeOutput}
	\end{CodeChunk}
	
\begin{figure}[h]
		\centering
		\includegraphics[width=0.7\textwidth]{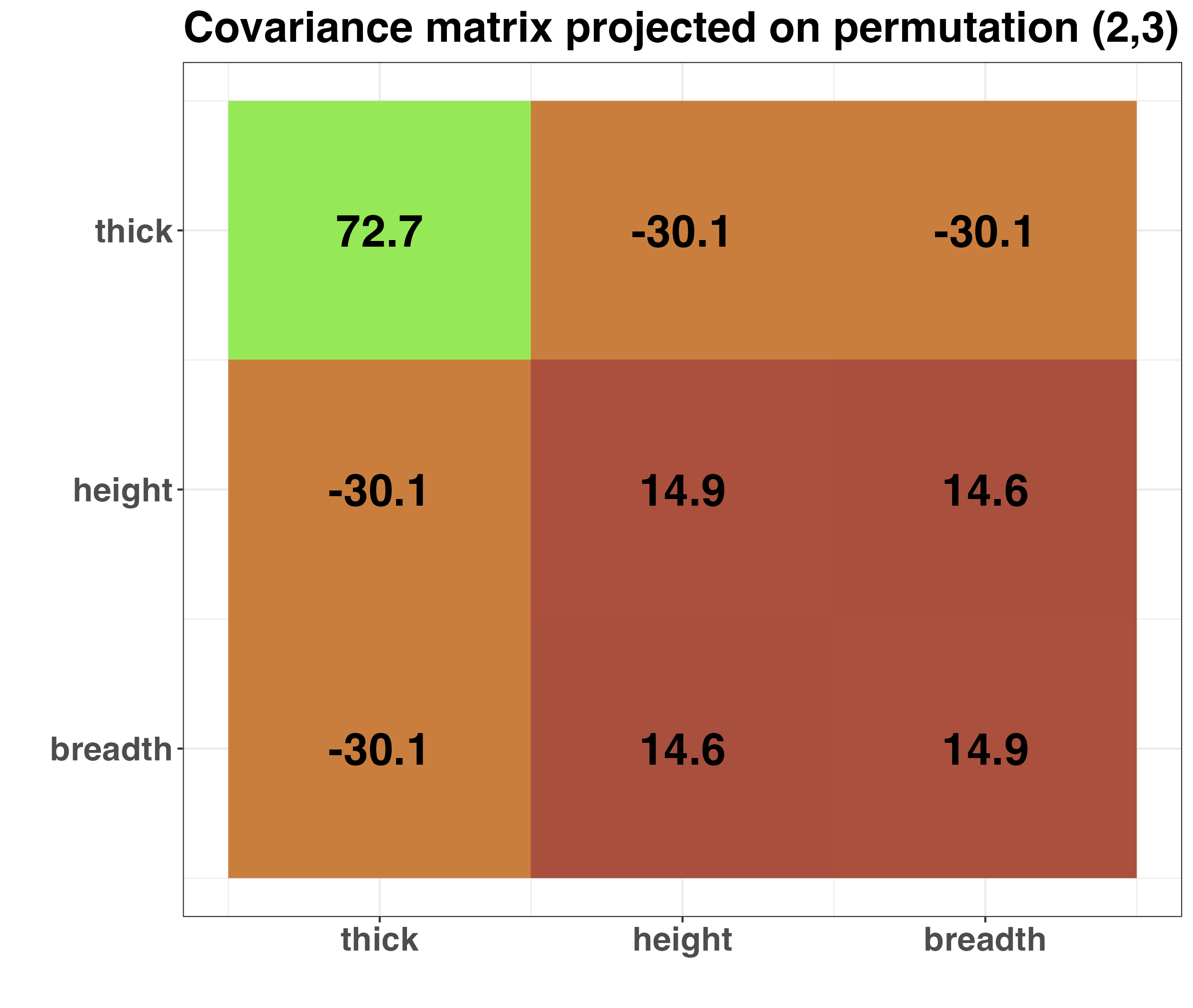}
		\caption{\small Estimator of the covariance matrix of the \code{oddbooks} dataset after the projection on the permutational symmetry group (2,3). We can see that now similar colors from Figure~\ref{fig:books_id} are exactly equal.}
		\label{fig:books_map}
	\end{figure}

	We see that the a posteriori distribution is maximized by a permutation \code{(2,3)}. The MLE of the covariance matrix in the model invariant under the permutation \code{(2,3)} is presented in Figure~\ref{fig:books_map}.
	
	\newpage
	\subsection{Motivation behind permutation symmetries} \label{sec:motivation}
	
	We argue that it is natural to expect certain symmetries in various applications, which strengthens the need for tools to investigate permutation symmetry within the data.
	
	For example, there are natural symmetries  in the data from gene expression. Specifically, the expression of a given gene is triggered by the binding of transcription factors to gene transcription factor binding sites. Transcription factors are proteins produced by other genes, often referred to as regulatory genes. Within the gene network, it is common for multiple genes to be triggered by the same regulatory genes, suggesting that their relative expressions depend on the abundance of the regulatory proteins (i.e., gene expressions) in a similar manner \cite{GIKM}. Extracting permutation symmetries can be utilized to identify genes with similar functions or groups of genes with similar interactions or regulatory mechanisms. This approach is particularly useful in unraveling the structures of gene regulatory networks \cite{KE20}.
	
	Furthermore, in examples of social networks, such as those influenced by geographical or social group clusters, additional symmetries must be taken into account, as mentioned in \cite{GM15}. In the study of the human brain's dynamics, it is believed that the left and right hemispheres possess a natural symmetric structure \cite{RRL21}.
	
	The discovery of hidden symmetries can greatly contribute to understanding complex mechanisms. Extracting patterns from gene expression profiles can offer valuable insights into gene function and regulatory systems \cite{TH02}. Clustering genes based on their expression profiles can aid in predicting the functions of gene products with unknown purposes and identifying sets of genes regulated by the same mechanism.
	
	\subsection{Arbitrary permutation symmetries vs cyclic permutation symmetries}\label{sec:vs}
	
	As observed in \cite{GIKM}, performing model selection within an arbitrary permutation subgroup is a highly challenging task. This difficulty arises not only due to theoretical reasons but also because of computational complexity issues arising when $p$ is large. Informally speaking, finding the parameters of an arbitrary permutation group becomes virtually impossible for large values of $p$. In \cite{GIKM}, a general model was developed; however, it was specifically applied to cyclic subgroups. Such subgroups are generated by a single permutation, and by restricting the analysis to them, efficient methods can be devised to conduct the model selection procedure. All the technical details regarding these methods will be presented in the subsequent sections.
	
	Furthermore, we argue that cyclic subgroups form a sufficiently rich family, as mentioned in \cite[Section 4.1]{GIKM}. Since these subgroups correspond to simpler symmetries, they are also more easily interpretable. Although our procedure exclusively explores cyclic subgroups, it can still provide valuable information even when the true subgroup is not cyclic, as discussed in \cite[Section 3.3]{GIKM_SM}. In fact, if the posterior probabilities (which are calculated with \pkg{gips}) are high for multiple groups, it is reasonable to expect that the data will exhibit invariance under the group containing those subgroups. We present a simple example in the Appendix \ref{app:general}.
	
	\newpage
	\section{Methodological background} \label{sec:models}

	After providing an informal introduction, let us proceed to define the key concepts and present the theory behind the \pkg{gips} package in a formal manner. Definitions in this section are accompanied with code in \pkg{gips} package. The running example is for $p=5$ and $n=10$. A standard PC can execute all the code in this section within $20$ seconds (except for the final chunk of code in Section \ref{sec:search} which runs for $2$ minutes).
	
	\begin{CodeChunk}
		\begin{CodeInput}
R> p <- 5; n <- 10
		\end{CodeInput}
	\end{CodeChunk}
	
	\subsection{Permutations} 
	
	Fix $p\in\{1,2,\ldots\}$. Let $\mathfrak{S}_p$ denote the symmetric group,  the set of all permutations on the set $V=\{1,\ldots,p\}$, with function composition as the group operation.
	
	Each permutation $\sigma\in\mathfrak{S}_p$ can be represented in a cyclic form. For example, if $\sigma$ maps $1$ to $2$, $2$ to $1$, and leaves $3$ unchanged, then we can express $\sigma$ as $(1,2)(3)$. It is sometimes convenient to exclude cycles of length $1$ from this representation. The identity permutation is denoted as $\mathrm{id}$ or $()$. The number of cycles denoted as $C_\sigma$, remains the same across different cyclic representations of $\sigma$. It is important to note that cycles of length $1$ are included when calculating $C_\sigma$.
	
	We say that a permutation subgroup $\Gamma\subset\mathfrak{S}_p$ is cyclic if $\Gamma=\{\sigma, \sigma^2,\ldots,\sigma^N\}=:\left<\sigma\right>$ for some $\sigma\in\mathfrak{S}_p$, where $N$ is the smallest positive integer such that $\sigma^N=\mathrm{id}$. Then, $N$ is the order of the subgroup $\Gamma$. 
	If $p_i$ denotes the length of the $i$th cycle in a cyclic decomposition of $\sigma\in\mathfrak{S}_p$, then $N$ is equal to the least common multiple of $p_1,p_2, \ldots, p_{C_\sigma}$. 
	
	If $\Gamma=\left<\sigma\right>$, then we say that $\sigma$ is a generator of $\Gamma$. It is worth noting that a cyclic subgroup may have several generators. Specifically, $\left<\sigma\right>=\langle\sigma^k\rangle$ for all $k=1,\ldots,N-1$, where $k$ is coprime with $N$. We identify each cyclic permutation subgroup by its generator, which is the smallest permutation according to lexicographic order. For further topics on permutation groups, readers may refer to \cite{Alg}.

	\subsection{Permutation symmetry}\label{sec:permSymmetry}
	
	Let $\Gamma$ be an arbitrary subgroup of $\mathfrak{S}_p$. We say that the distribution of $Z=(Z_i)_{i\in V}$ is invariant under a subgroup $\Gamma$ if $Z$ has the same distribution as $(Z_{\sigma(i)})_{i\in V}$ for all $\sigma\in\Gamma$. If $Z$ is a multivariate random variable following a centered Gaussian distribution $\mathrm{N}_p(0,\Sigma)$, then this invariance property can be expressed as a condition on the covariance matrix. Specifically, the distribution of $Z$ is invariant under $\Gamma$ if and only if for all $i,j\in V$:
	\begin{align}\label{eq:sigma}
		\Sigma_{ij}=\Sigma_{\sigma(i)\sigma(j)}\quad\text{for all}\quad \sigma\in\Gamma.
	\end{align}
	
\begin{figure}[h!]
		\centering
		\includegraphics[width=0.45\textwidth]{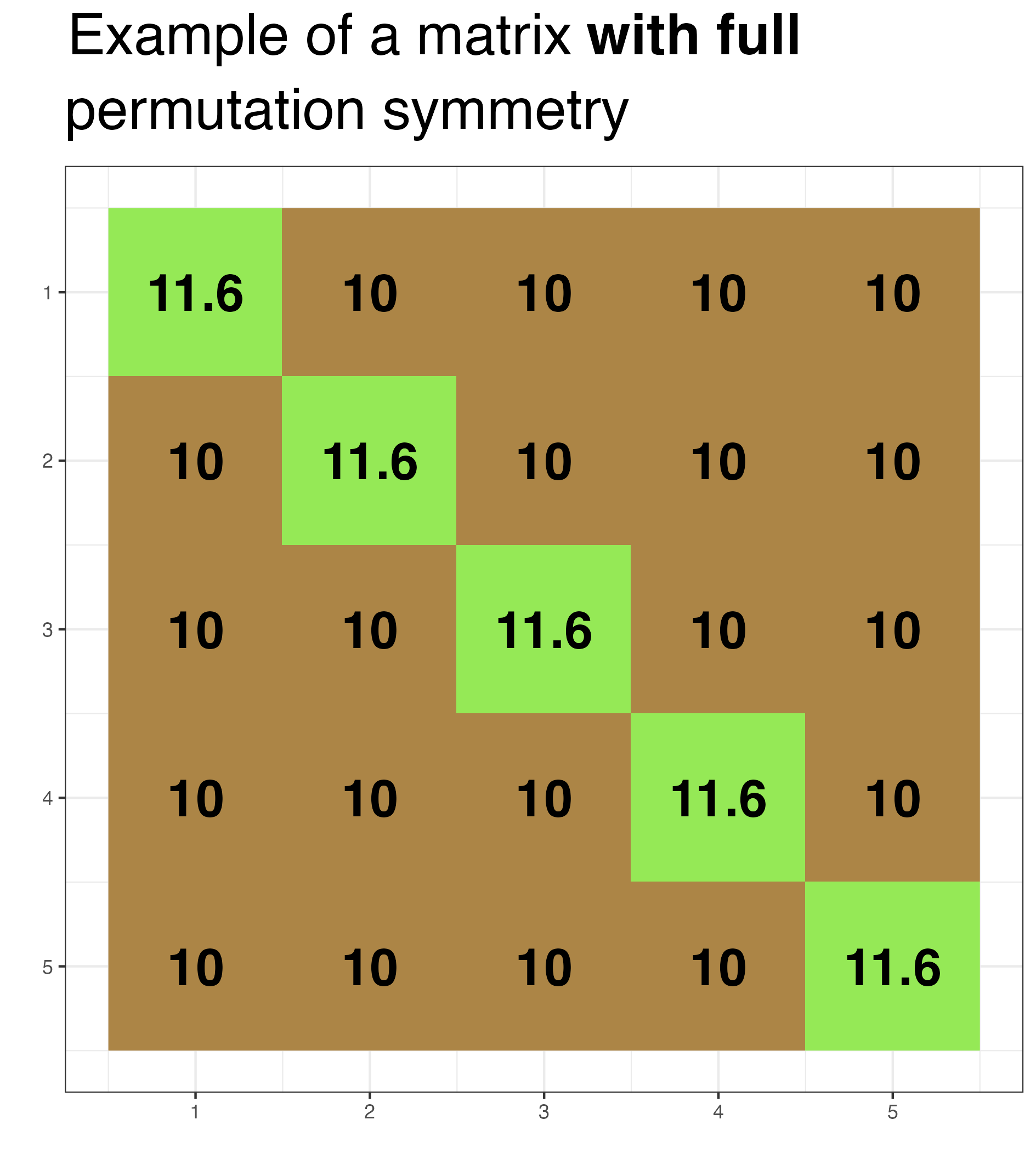}\quad
		\includegraphics[width=0.45\textwidth]{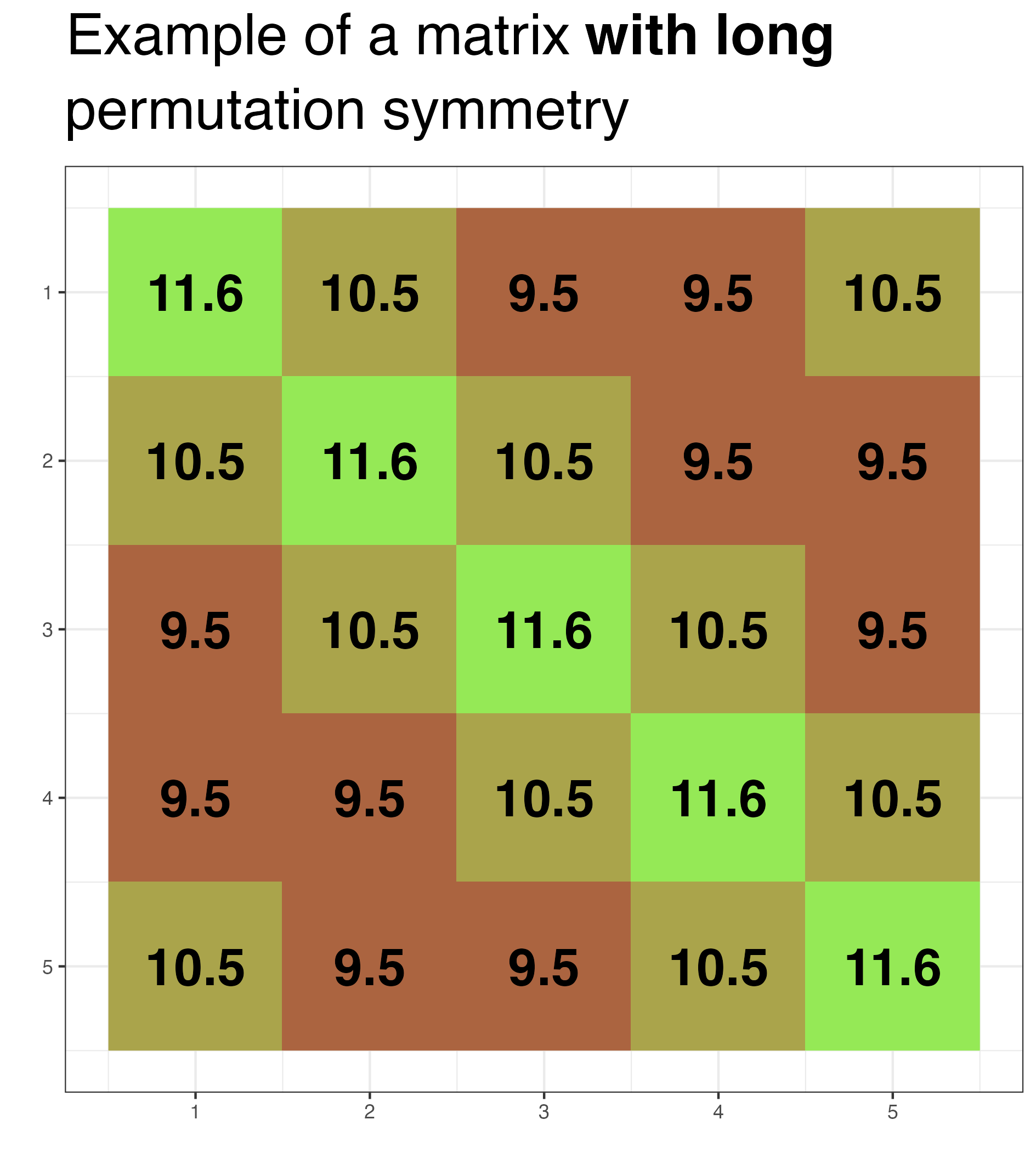}\\
		\includegraphics[width=0.45\textwidth]{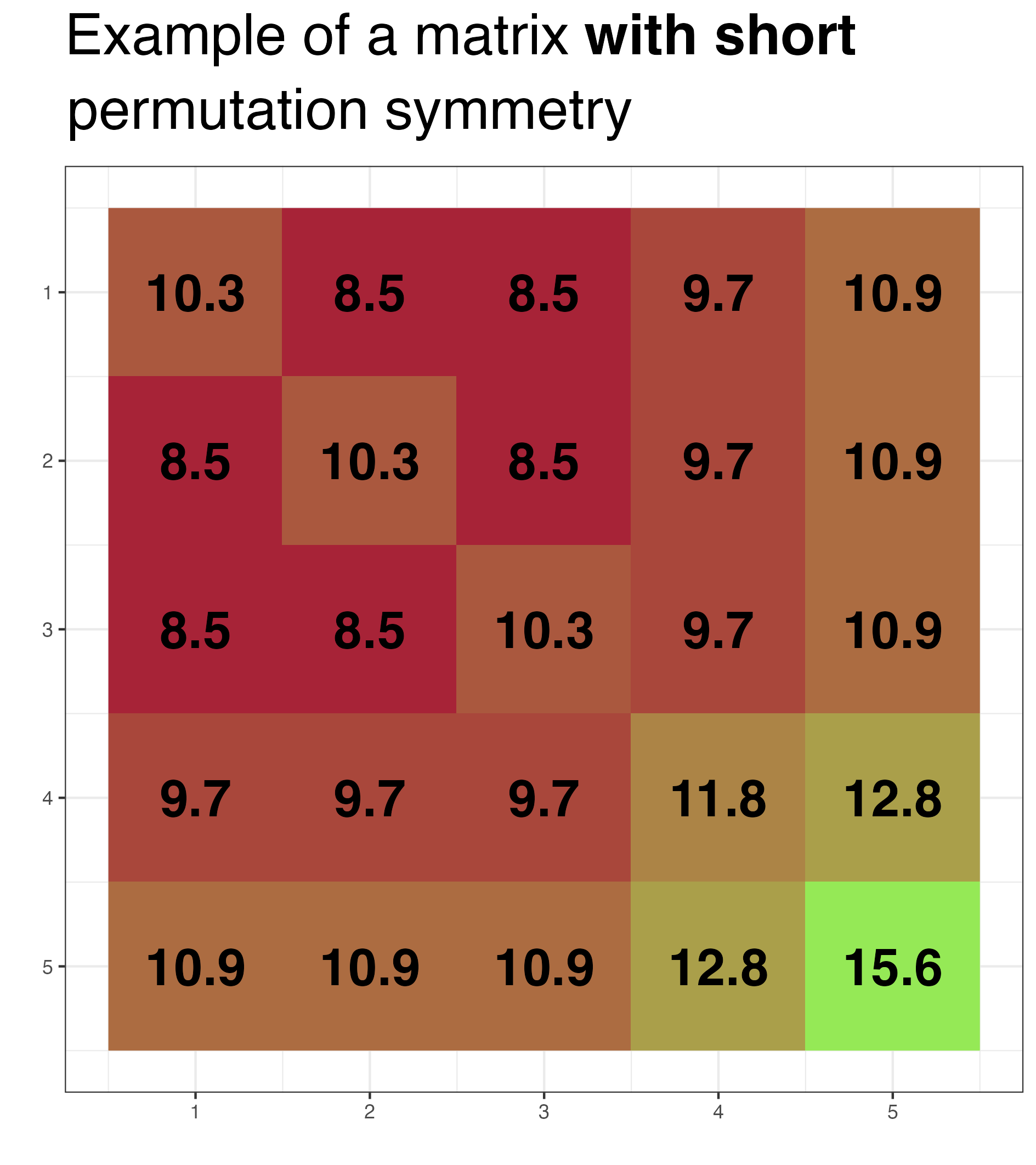}\quad
		\includegraphics[width=0.45\textwidth]{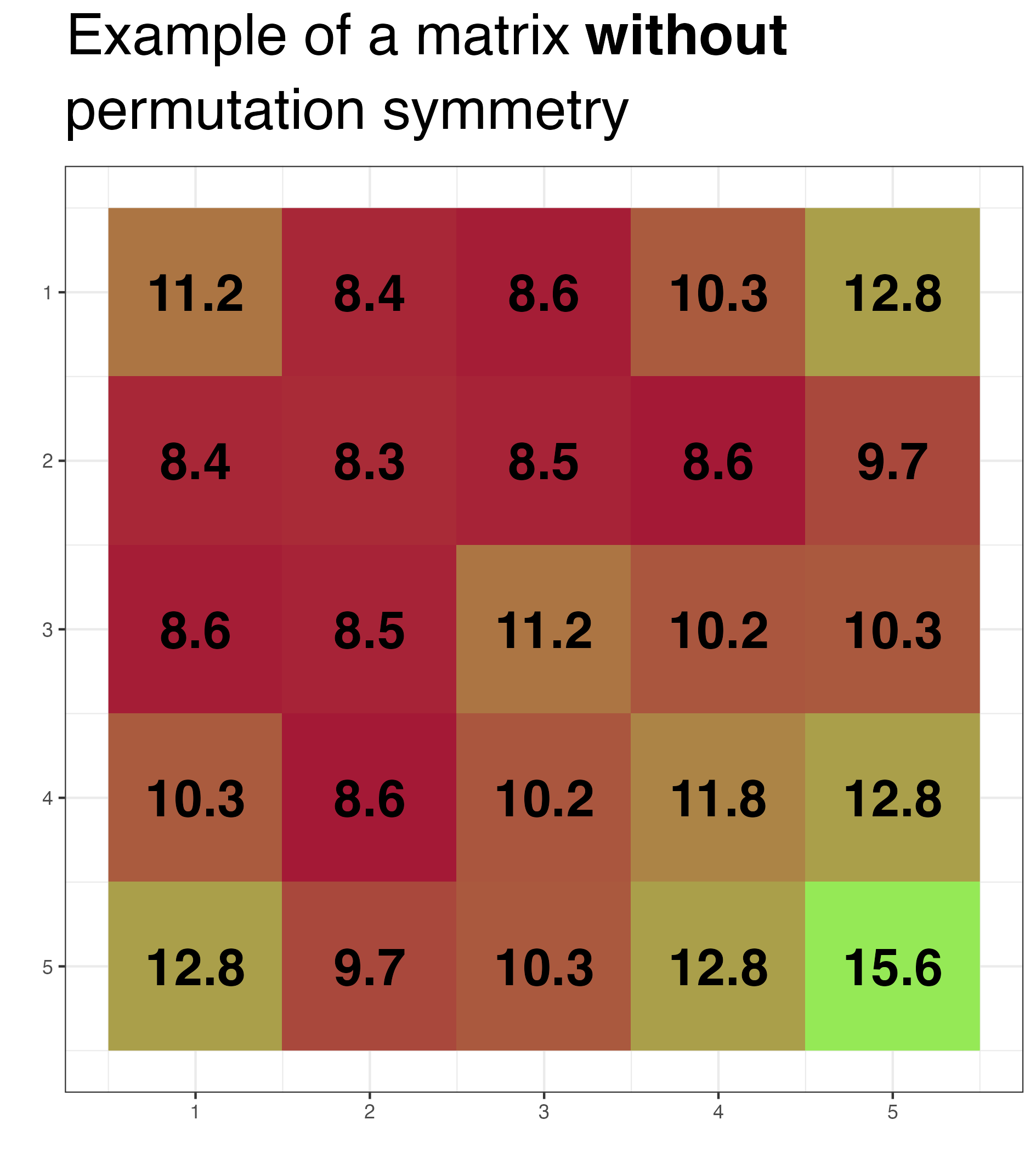}
		\caption{\small Examples of covariance matrices with symmetries: full $\Gamma=\mathfrak{S}_p$ permutation symmetry (top left), long permutation symmetry $\Gamma=\left<(12345)\right>$ (top right), short permutation symmetry $\Gamma=\left<(123)\right>$ (bottom left), no permutation symmetry (bottom right). They all are projections of the same matrix on corresponding colored spaces.}
		\label{fig:example_symmetries}
	\end{figure}

	When $\Gamma=\mathfrak{S}_p$, the above conditions imply that all diagonal entries of $\Sigma$ are the same, and similarly, the off-diagonal entries are the same (see the top left panel of Figure~\ref{fig:example_symmetries}). On the other hand, if $\Gamma$ is the trivial subgroup, i.e., $\Gamma=\{\mathrm{id}\}$, then (\ref{eq:sigma}) does not impose any restrictions on the entries of $\Sigma$. If $\Gamma$ is non-trivial, the sample size $n$ required for the MLE to exist is lower than $p$, as discussed in Section \ref{sec:MLE}.
	
	Let $\mathrm{Sym}(p;\mathbb{R})$ and $\mathrm{Sym}^+(p;\mathbb{R})$ denote the space of $p\times p$ symmetric matrices and the corresponding cone of positive definite matrices, respectively. For a subgroup $\Gamma \subset \mathfrak{S}_p$, we define the colored space as the space of symmetric matrices invariant under $\Gamma$:
	\[
	\mathcal{Z}_{\Gamma} 
	:= \{S \in \mathrm{Sym}(p;\mathbb{R})\colon S_{ij} = S_{\sigma(i)\sigma(j)} \mbox{ for all }\sigma \in \Gamma\},
	\]
	We also define the colored cone of positive definite matrices valued in $\mathcal{Z}_{\Gamma}$ as:
	\[
	\mathcal{P}_{\Gamma} := \mathcal{Z}_{\Gamma} \cap \mathrm{Sym}^+(p;\mathbb{R}).
	\]
	The set $\mathcal{P}_\Gamma$ contains all possible covariance matrices of Gaussian vectors invariant under subgroup $\Gamma$. 
	The dimension of the space $\mathcal{Z}_\Gamma$ corresponds to the number of free parameters in the covariance matrix. The dependence structure of a Gaussian vector is fully described by the covariance matrix $\Sigma$. When certain entries of $\Sigma$ are identical, we refer to them as having the same color. There are colors that correspond to equalities among the diagonal elements of $\Sigma$, and there are independent colors that correspond to equalities among the off-diagonal elements of $\Sigma$. Thus, in the context of colored models, $\dim(\mathcal{Z}_\Gamma)$ can be interpreted as the number of distinct colors.
	
	In \pkg{gips}, we can easily find the number of free parameters in the model invariant under a cyclic subgroup as follows (\code{S} is a matrix from the bottom right of Figure~\ref{fig:example_symmetries}):
	\begin{CodeChunk}
		\begin{CodeInput}
R> g <- gips(S, n, perm = "(12345)", was_mean_estimated = FALSE)
R> summary(g)$n_parameters
		\end{CodeInput}
		\begin{CodeOutput}
3
		\end{CodeOutput}
	\end{CodeChunk}
	
	Notice there were exactly $3$ different numbers in the matrix from the top right of Figure~\ref{fig:example_symmetries}.
	
	It is important to note that the mapping $\Gamma\mapsto\mathcal{Z}_\Gamma$ is not one-to-one. In particular, for $p=3$, we have $\mathcal{Z}_{\left<(1,2,3)\right>}=\mathcal{Z}_{\mathfrak{S}_3}$.
	A notable property of cyclic subgroups is that they correspond to different colored spaces. More precisely, if $\mathcal{Z}_{\left<\sigma\right>}=\mathcal{Z}_{\left<\sigma'\right>}$ for some $\sigma,\sigma'\in\mathfrak{S}_p$, then $\left<\sigma\right>=\left<\sigma'\right>$ \cite[Lemma 15]{GIKM}.

	\subsection{The MLE in the Gaussian model invariant under permutation symmetry}\label{sec:MLE}
	Let $Z^{(1)},\ldots,Z^{(n)}$ be an independent and identically distributed (i.i.d.) sample from $\mathrm{N}_p(0, \Sigma)$. The presence of equality restrictions in (\ref{eq:sigma}) reduces the number of parameters to estimate in permutation invariant models. Consequently, the sample size required for the MLE of $\Sigma$ to exist is lower than $p$ for non-trivial subgroup $\Gamma\subset\mathfrak{S}_p$. Assuming $\Sigma \in \mathcal{P}_\Gamma$, where $\Gamma=\left<\sigma\right>$ is a cyclic subgroup, \cite[Corollary 12]{GIKM} establishes that the MLE of $\Sigma$ exists if and only if
	\begin{align}\label{eq:n0}
		n \geq n_0 := C_\sigma.
	\end{align}
	In particular, when $\sigma = \mathrm{id}$, no restrictions are imposed on $\Sigma$, and we recover the well-known condition that the sample size $n$ must be greater than or equal to the number of variables $p=C_{\mathrm{id}}$. However, if $\sigma$ consists of a single cycle, i.e., $C_\sigma = 1$, the MLE always exists. This remarkable observation is crucial in high-dimensional settings. 
	
	In \pkg{gips}, we can compute $n_0$ as follows:
	\begin{CodeChunk}
		\begin{CodeInput}
R> g <- gips(S, n, perm = "(12345)", was_mean_estimated = FALSE)
R> summary(g)$n0
		\end{CodeInput}
		\begin{CodeOutput}
1
		\end{CodeOutput}
		\begin{CodeInput}
R> g <- gips(S, n, perm = "()", was_mean_estimated = FALSE)
R> summary(g)$n0
		\end{CodeInput}
		\begin{CodeOutput}
5
		\end{CodeOutput}
	\end{CodeChunk}
	
	If (\ref{eq:n0}) is satisfied, the MLE of $\Sigma$ is given by
	\[
	\hat{\Sigma} =  \pi_{\Gamma}\left(\frac{1}{n}\sum_{i=1}^n Z^{(i)}\cdot (Z^{(i)})^\top\right),
	\]
	where $\pi_\Gamma$ denotes the orthogonal projection onto the colored space $\mathcal{Z}_\Gamma$. It is is defined as
	\begin{align*}
		\pi_\Gamma(X) = \frac{1}{\#\Gamma} \sum_{\sigma\in\Gamma} \sigma\cdot X\cdot \sigma^\top,
	\end{align*}
	where each permutation $\sigma$ is identified with its corresponding permutation matrix. The resulting matrix $\pi_\Gamma(X)$ is often referred to as the regularized matrix since the mapping averages the entries of $X$ that correspond to the same orbits of $\Gamma$:
	for $\{i,j\}\subset V$ define its $\Gamma$-orbit by $O_{ij}^\Gamma=\{ \{\sigma(i),\sigma(j)\}\colon \sigma\in \Gamma\}$. Then, for any $\{u,v\}\in O_{ij}^\Gamma$ one has
	\begin{align}\label{pig}
		\pi_\Gamma(X)_{uv} = \frac{1}{\#O_{ij}^\Gamma} \sum_{\{k,l\}\in O_{ij}^\Gamma} X_{kl}.
	\end{align}
	
	In \pkg{gips}, the projection $\pi_{\left<\texttt{perm}\right>}(\texttt{S})$ of a matrix \code{S} onto $\mathcal{Z}_{\left<\texttt{perm}\right>}$ is calculated as follows:
	\begin{CodeChunk}
		\begin{CodeInput}
R> S_projected <- project_matrix(S, perm)
		\end{CodeInput}
	\end{CodeChunk}
	where \code{perm} can be the permutation of a form \code{"(12345)"}, or an object of a \code{`gips`} class.
	
	\subsection{Bayesian model selection procedure}\label{sec:BM}
	
	Now we shift our focus to methods aimed at discovering permutation symmetries in the data. The model introduced in \cite{GIKM} is considered. In this model, the multivariate Gaussian sample $Z^{(1)},\ldots, Z^{(n)}$ given $\{K=k, \Gamma=c\}$ consists of i.i.d. $\mathrm{N}_p(0,k^{-1})$ random vectors.
	
	Let $\Gamma$ be a discrete random variable uniformly distributed over the set $\mathcal{C}:=\{\left<\sigma\right>\colon\sigma\in\mathfrak{S}_p\}$ of cyclic subgroups of $\mathfrak{S}_p$. It is assumed that $K$ given $\{\Gamma=c\}$  follows the Diaconis-Ylvisaker conjugate prior \citep{DiaconisYlvisaker} distribution, defined by its density
	\[
	f_{K|\Gamma=c}(k)=\frac1{I_{c} (\delta,D)} {\mathrm{Det}(k)^{(\delta-2)/2} e^{- \tfrac12 \mathrm{Tr}[D\cdot k]} }
	{\bf 1}_{\mathcal{P}_{c}}(k),
	\]
	where $\delta>1$ and $D\in \mathcal{P}_{c}$ are the hyperparameters, and $I_c(\delta,D)$ is the normalizing constant. 
	
	It was derived in \cite{GIKM} that the posterior probability is proportional to
	\begin{align}\label{eq:defpp}
		\Prob\left(\Gamma=c|Z^{(1)},\ldots,Z^{(n)}\right) \propto \frac{ I_{c}(\delta + n, D+U)}{I_{c}(\delta,D)},\qquad c\in\mathcal{C},
	\end{align}
	where $U=\sum_{i=1}^n Z^{(i)}\cdot (Z^{(i)})^\top$. In order to utilize quotient~(\ref{eq:defpp}), it is necessary to calculate or approximate the ratios of the normalizing constants. An efficient method for calculating these constants for cyclic subgroups was introduced in \cite{GIKM}. This method relies on the block decomposition of the colored space $\mathcal{Z}_\Gamma$ and is implemented in the \pkg{gips} package. Further technical details are provided in Appendix \ref{app:technical}.
	
	In \pkg{gips}, one can calculate the quotient on the right-hand side of quotient~(\ref{eq:defpp}) for permutation $c=\langle$\code{(12345)}$\rangle$ as follows:
	\begin{CodeChunk}
		\begin{CodeInput}
R> g <- gips(S, n, perm = "(12345)", was_mean_estimated = FALSE)
R> exp(log_posteriori_of_gips(g))
		\end{CodeInput}
		\begin{CodeOutput}
4.586821e-27
		\end{CodeOutput}
	\end{CodeChunk}
	
	This is a very small number, but keep in mind that the posteriori probability of a subgroup $c$ is proportional to this quantity (not equal). One can compare it with other subgroups to get interpretable result (under our Bayesian setting):
	\begin{CodeChunk}
		\begin{CodeInput}
R> compare_posteriories_of_perms(g, "(123)")
		\end{CodeInput}
		\begin{CodeOutput}
The permutation (1,2,3,4,5) is 22.827 times more likely
than the (1,2,3) permutation.
		\end{CodeOutput}
	\end{CodeChunk}
	
	Following the Bayesian paradigm, we work with the maximum a posteriori (MAP) estimator, which corresponds to the cyclic subgroup with the highest posterior probability, i.e., this estimator is defined as
	\begin{align}\label{eq:gammahat}
		\hat{\Gamma} = \operatorname{arg\,max}_{c\in\mathcal{C}} \Prob\left(\Gamma=c|Z^{(1)},\ldots,Z^{(n)}\right).
	\end{align}
	
	While the choice of hyperparameters is not scale invariant, it is a common practice in similar models to set $\delta=3$ and $D=I_p$, \cite{MassamBayesian}. The parameter $\delta=3$ serves  as the default parameter in our method, but we decided to set $D=\frac{\mathrm{tr}(S)}{p}\cdot I_p$ as default and justify our choice in Section \ref{sec:standardizing}. In \pkg{gips}, one can pass the desired values of these parameters via \code{delta} and \code{D_matrix} arguments in \code{gips()} function.
	
	In Section \ref{sec:hyper_influence}, we considered the influence of $\delta$ and $d$ in $D = d \cdot I_p$. The role of the $d$ parameter turns out to be quite similar to the role of the tuning parameters $\lambda$ in LASSO (least absolute shrinkage and selection operator) methods. To conclude, smaller values of $d$ tend to favor large symmetries. Therefore, through such exploratory analysis, users can adjust the parameter $d$ to achieve a model that aligns most meaningfully with their preferences and requirements.

	\subsection{Searching for a MAP estimator}\label{sec:search}
	
	The quotient~(\ref{eq:defpp}) enables the numerical evaluation of how well a given permutation symmetry (specifically, a cyclic group generated by a permutation) fits the data. Finding a cyclic subgroup with a high evaluation score is a challenging task for large values of $p$ due to the vast size of the space of potential permutation symmetries.
	
	Recall that $\mathcal{C}$ is the space of cyclic subgroups of $\mathfrak{S}_p$.
	For small values of $p$ (in our investigation, up to $9$), it is possible to compute the posterior quotients~(\ref{eq:defpp}) for all $c\in\mathcal{C}$ and determine $\hat{\Gamma}$ from (\ref{eq:gammahat}) using exact calculations, i.e., for $c\in\mathcal{C}$ we have
	\begin{align}\label{eq:exact}
		\Prob\left(\Gamma=c|Z^{(1)},\ldots,Z^{(n)}\right) = \frac{ I_{c}(\delta + n, D+U)}{{I_{c}(\delta,D)} \sum_{s\in\mathcal{C}} \frac{ I_{s}(\delta + n, D+U)}{I_{s}(\delta,D)} }.
	\end{align}

	However, the cardinality of $\mathcal{C}$ grows super-exponentially with $p$. Specifically, for $p=150$, the cardinality of $\mathcal{C}$ is approximately $10^{250}$. This makes it computationally infeasible to calculate the quotients~(\ref{eq:defpp}) for all $c\in\mathcal{C}$.
	
	To address this challenge, we propose the use of a Monte Carlo Markov chain method. We define an irreducible Markov chain $(\sigma_t)_{t}$ that traverses an even larger space, $\mathfrak{S}_p$, and apply the Metropolis-Hastings algorithm to obtain preliminary estimates of the posterior probabilities. Subsequently, taking into account the fact that some permutations generate the same cyclic subgroup, we derive the estimates of the posterior probabilities using equation (\ref{eq:almost_sure_to_exact}) below. Through the ergodic theorem, the Metropolis-Hastings algorithm provides statistical guarantees that the estimates will converge to the true values as the number of iterations tends to infinity.

	A transposition is a permutation that swaps two elements while leaving other elements unchanged. In other words, each transposition is in the form $(i,j)$ for some $i,j\in V$ where $i\neq j$. Let $\mathcal{T}$ denote the set of all transpositions.
	\begin{algorithm}
		\caption{Metropolis-Hastings algorithm}\label{alg:cap}
		\begin{algorithmic}
			\State Let $T\in\mathbb{N}$.
			\State Let $\sigma_0$ be an arbitrary permutation from $\mathfrak{S}_p$.
			\For{$t=1,2,\ldots,T$}
			\State Sample $x_t$ uniformly from the set $\mathcal{T}$ and set $\sigma^\prime =\sigma_{t-1}\circ x_t$.
			\State Accept the move $\sigma_t = \sigma^\prime$ with probability
			\[
			\min\left\{ 1, \frac{I_{\left<\sigma^\prime\right>}(\delta+n,D+U)\,\cdot\, I_{\left<\sigma_{t-1}\right>}(\delta,D) }{I_{\left<\sigma^\prime\right>}(\delta,D)\,\cdot\, I_{\left<\sigma_{t-1}\right>}(\delta+n,D+U)  } 
			\right\}.
			\]
			\State If the move is rejected, set $\sigma_t=\sigma_{t-1}$.
			\EndFor
		\end{algorithmic}
	\end{algorithm}
	
	The algorithm produces a sequence of permutations $(\sigma_t)_{t=1}^T$, and we construct a corresponding sequence of cyclic subgroups $(\left<\sigma_t\right>)_{t=1}^T$. The MAP estimator, which corresponds to the cyclic group with the highest posterior probability, is given by
	\[
	\hat{\Gamma} = \operatorname{arg\,max}_{c\in(\left<\sigma_t\right>)_{t=1}^T} \Prob\left(\Gamma=c|Z^{(1)},\ldots,Z^{(n)}\right)=\operatorname{arg\,max}_{c\in(\left<\sigma_t\right>)_{t=1}^T} \frac{I_c(\delta+n,D+U)}{I_c(\delta,D)}.
	\]
	where the maximum is taken over all permutations visited by the Markov chain constructed in the algorithm.
	
	The Algorithm \ref{alg:cap} is implemented in \code{find_MAP()} function with parameter \code{optimizer = "MH"}:
	\begin{CodeChunk}
		\begin{CodeInput}
R> g <- gips(S, n, was_mean_estimated = FALSE)
R> g_MAP_MH_25 <- find_MAP(g, max_iter = 25, optimizer = "MH")
R> g_MAP_MH_25
		\end{CodeInput}
		\begin{CodeOutput}
The permutation (1,2,3,5):
 - was found after 25 posteriori calculations;
 - is 5.149 times more likely than the () permutation.
		\end{CodeOutput}
	\end{CodeChunk}
	
	The algorithm in $25$ steps found a quite long permutation. Always keep in mind the Algorithm \ref{alg:cap} is only an approximate one. However, if one wants to have the true MAP, a brute-force search can be applied:
	\begin{CodeChunk}
		\begin{CodeInput}
R> g_MAP_BF <- find_MAP(g, optimizer = "BF")
R> g_MAP_BF
		\end{CodeInput}
		\begin{CodeOutput}
The permutation (1,2,3,4,5):
 - was found after 67 posteriori calculations;
 - is 33.743 times more likely than the () permutation.
		\end{CodeOutput}
		\begin{CodeInput}
R> compare_posteriories_of_perms(g_MAP_BF, g_MAP_MH_25)
		\end{CodeInput}
		\begin{CodeOutput}
The permutation (1,2,3,4,5) is 6.553 times more likely
than the (1,2,3,5) permutation.
		\end{CodeOutput}
	\end{CodeChunk}
	
	If one is interested in estimating $\Prob\left(\Gamma=c|Z^{(1)},\ldots,Z^{(n)}\right)$ for an arbitrary $c\in\mathcal{C}$, the following approach can be used. For a permutation subgroup $\Gamma$, let $\varphi(\Gamma)=\Phi(\#\Gamma)$, where $\Phi$ is the Euler totient function, i.e., $\Phi(n)=\#\{k\in \{1,\ldots,n\}\colon k\mbox{ and }n\mbox{ are coprime}\}$. In \cite[Theorem 16]{GIKM}, it is shown that as $T\to\infty$ and for $c\in\mathcal{C}$,
	\begin{align}\label{eq:almost_sure_to_exact}
		\hat{\pi}_c:=\frac{\sum_{t=1}^T {\bf 1}(\left<\sigma_t\right>=c) }{\varphi(\left<c\right>)  \sum_{t'=1}^T  1/\varphi(\left<\sigma_{t'}\right>)}\stackrel{a.s.}{\longrightarrow} \Prob\left(\Gamma=c|Z^{(1)},\ldots,Z^{(n)}\right).
	\end{align}
	In practice, $\hat{\pi}_c$ serves as an approximation to $\Prob\left(\Gamma=c|Z^{(1)},\ldots,Z^{(n)}\right)$ for large $T$. 
	
	By default, \pkg{gips} does not save all the computed permutations, but only the best one. One can set the flag \code{save_all_perms = TRUE} to get the desired exact distribution:
	\begin{CodeChunk}
		\begin{CodeInput}
R> g_MAP_BF_with_probs <- find_MAP(g,
+    optimizer = "BF",
+    save_all_perms = TRUE, return_probabilities = TRUE
+  )
R> head(get_probabilities_from_gips(g_MAP_BF_with_probs), 10)
		\end{CodeInput}
		\begin{CodeOutput}
 (1,2,3,4,5)   (1,4)(2,3)   (1,2)(3,5)   (1,3)(4,5)  (1,2,4,5,3)   (1,5)(2,4) 
  0.13478976   0.05532886   0.05487531   0.04410568   0.03076733   0.02969751 
   (1,2,5,3)    (1,2,3,4) (1,2,4)(3,5)    (1,3,4,5) 
  0.02928655   0.02692967   0.02638740   0.02501291 
		\end{CodeOutput}
	\end{CodeChunk}
	
	If one wants to estimate the distribution, e.g., when $p$ is too large to search through entire space, one can do exactly the same with the \code{optimizer = "MH"}:
	\begin{CodeChunk}
		\begin{CodeInput}
R> g_MAP_MH_20000 <- find_MAP(g,
+    optimizer = "MH", max_iter = 20000,
+    save_all_perms = TRUE, return_probabilities = TRUE
+  )
R> head(get_probabilities_from_gips(g_MAP_MH_20000), 10)
		\end{CodeInput}
		\begin{CodeOutput}
 (1,2,3,4,5)   (1,4)(2,3)   (1,2)(3,5)   (1,3)(4,5)   (1,5)(2,4)    (1,2,3,4) 
  0.13832762   0.06678977   0.05718808   0.04167766   0.03102084   0.02970193 
   (1,2,5,3)  (1,2,4,5,3) (1,2,4)(3,5)    (1,3,4,5) 
  0.02964917   0.02930625   0.02653653   0.02495384 
		\end{CodeOutput}
	\end{CodeChunk}
	
	We can observe that the estimated probabilities are similar to the true ones. 
	However, please note that the above code is intended solely to demonstrate convergence, and in practical scenarios, it is unreasonable to execute the function \code{find_MAP(optimizer = "MH", max_iter = my_max_iter)} for \code{my_max_iter} $> p!$ due to the faster and exact brute-force \code{find_MAP(optimizer = "BF")}. The \pkg{gips} package will show a warning in such a case.

	\subsection{Scaling, centering and standardizing data}\label{sec:standardizing}
	We want to emphasize that the considered model of permutation symmetries is not scale-invariant in the following sense: if $Z$ is invariant under a subgroup, a random vector $\mathrm{diag}(\alpha)\cdot Z$, where $\alpha\in\mathbb{R}^p$, is generally not invariant under any permutation subgroup. Therefore, it is recommended to apply our procedure to data that have comparable scales and keep all variables in the same units. That is the reason for the division of \code{height} by $\sqrt{2}$ at the beginning of example \textit{Books dataset} from Section \ref{sec:toyexps}. 
	
	It is worth noting that there are many examples of such data, such as gene expression data, where measurements are on the same scale due to being results of experiments of the same type and measured using the same gauges. For further references, see e.g., \cite{Ge11, GM15, MassamBayesian, Hel20, QXNX21, RRL21}.
	
	However, our model is scale-invariant under common scaling, i.e., if $Z$ is invariant under $\Gamma$, then $\beta Z$ for any $\beta\in\mathbb{R}$ is also invariant under $\Gamma$.  Our practice shows that choosing $\beta$ in a way that $\beta Z$ has \emph{average} unit variance often produces good results. Such scaling can be equivalently accomplished by choosing the hyperparameter $D=\frac{\mathrm{tr}(S)}{p}\cdot I_p$, where $S$ is the empirical covariance matrix of $Z$. This is the default parameter for \code{D_matrix} in \code{gips()} function.
	
	While our Bayesian model is designed for a zero mean Gaussian sample, it can be easily extended to handle samples with arbitrary means. If $Z^{(1)},\ldots,Z^{(n)}$ is an i.i.d. sample from $\mathrm{N}_p(\mu,\Sigma)$, the user can center the data and take this into account by setting the parameter \code{was_mean_estimated = TRUE} in function \code{gips()}.
	
	In cases where the sample size $n$ is reasonably large, it is common to assume that the standardized normal sample (which follows a multivariate $t$-distribution) can be approximated by a Gaussian distribution. Therefore, for large $n$, one can standardize each variable and apply our model selection procedure to obtain reliable estimates. However, it is important to note that after standardization, the empirical covariance matrix will have a unit diagonal, which may favor cyclic subgroups whose generators consist of a single cycle, as they correspond to matrices with a constant diagonal. This is the reason why we do not recommend standardizing the data when the sample size $n$ is small.
	
	\section{Package structure and usage} \label{sec:illustrations}

	The package \pkg{gips} is available under the general public license (GPL$\geq$3) from the Comprehensive \proglang{R} Archive Network (CRAN) at \url{https://CRAN.R-project.org/package=gips}. \hbox{Its documentation} is available as \pkg{pkgdown} page at \url{https://przechoj.github.io/gips}.
	\hbox{\pkg{gips} can} be installed and loaded into the current \proglang{R} session using the following code:
	\begin{CodeChunk}
		\begin{CodeInput}
R> install.packages("gips")
R> library("gips")
		\end{CodeInput}
	\end{CodeChunk}
	
	\begin{figure}[h]
		\centering
		\includegraphics[width=\textwidth]{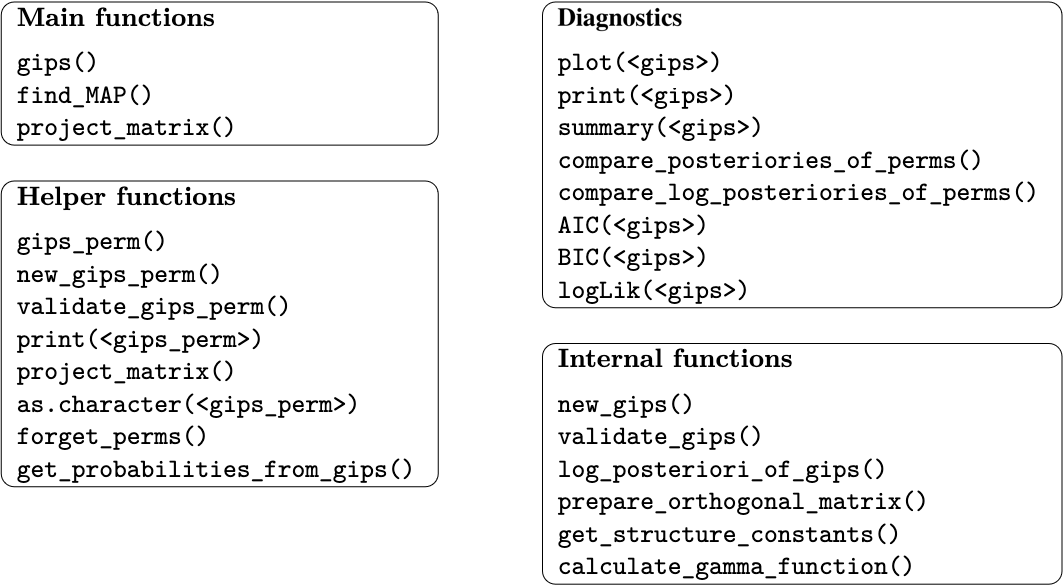}
		\caption{\small The structure of the \pkg{gips} package.}
		\label{fig:structure}
	\end{figure}
	
	The primary use case of the \pkg{gips} package is to find a permutation subgroup with the maximum a posteriori probability in the Bayesian model introduced in the previous sections and estimate the covariance matrix in the model invariant under this permutation subgroup. Representations and operations on permutations are performed using the \pkg{permutations} package. We decided to use this package due to its ease of transforming permutations and its compactness.
	
	We start the description of the main functions implemented in the proposed package. 
	The workflow in \pkg{gips} is as follows: first, use the \code{gips()} function to define an object of the \code{`gips`} class that contains all the necessary information for the model.
	\begin{CodeChunk}
		\begin{CodeInput}
R> g <- gips(S, number_of_observations, delta = 3, D_matrix = NULL, 
+            was_mean_estimated = TRUE, perm = "")
		\end{CodeInput}
	\end{CodeChunk}
	Let $Z$ be a data matrix of size \code{number_of_observations}$\times p$. The parameter \code{S} is the $p\times p$ empirical covariance matrix of $Z$ and \code{number_of_observations} is the corresponding sample size. 
	If one does not know the theoretical mean of the distribution data was sampled from, use \code{S = cov(Z)} and leave the flag \code{was_mean_estimated = TRUE} as default. If the theoretical mean is known to be $0$, use \code{S = (t(Z) \%*\% Z) / number_of_observations}, and set the flag \code{was_mean_estimated = FALSE},
	
	Parameters \code{delta} and \code{D_matrix} are the hyperparameters of our Bayesian model. The domains of these parameters are the following: \code{delta} $>1$ and \code{D_matrix} has to be $p\times p$ positive definite matrix. The default value of \code{D_matrix} is \code{mean(diag(S)) * diag(p)}. The last parameter, \code{perm}, is an optional permutation on $p$ elements. Can be of any form that the function \code{permutations::permutation()} can handle. This is the starting permutation for the Metropolis-Hastings.
	
	Next, use the \code{find_MAP()} function with an optimizer of your choice to find the permutation that provides the maximum a posteriori estimate. 
	
	\begin{CodeChunk}
		\begin{CodeInput}
R> find_MAP(g, max_iter = NA, optimizer = NA, show_progress_bar = TRUE,
+           save_all_perms = FALSE, return_probabilities = FALSE)
		\end{CodeInput}
	\end{CodeChunk}
	The first parameter, \code{g}, is the object of \code{`gips`} class. There are three optimizers implemented: 
	\begin{description}
		\item[\code{optimizer = "BF"}:] brute-force search (default and only intended for $p\leq 9$). 
		\item[\code{optimizer = "MH"}:] The Metropolis-Hasting algorithm (default and recommended for $p > 9$).
		\item[\code{optimizer = "HC"}:] The hill climbing algorithm.
	\end{description}
	The parameter \code{max\_iter} is the number of steps for the Metropolis-Hastings and the hill climbing algorithm. For the Metropolis-Hastings algorithm, it has to be finite and greater than $2$, while for hill climbing it can be also \code{Inf}. 
	
	The progress bar of the optimization process can be turned on and off by changing the boolean parameter \code{show_progress_bar}.
	
	To obtain the entire posterior distribution, the flag \code{return\_probabilities} has to be set to \code{TRUE}. This flag can only be provided when also \code{save_all_perms = TRUE}, which saves a list of all permutations that were visited during optimization.
	In the case of \code{optimizer = "BF"}, the exact posterior probabilities are calculated. For other optimizers, their estimates are calculated.
	One can access these probabilities using the function 
	\begin{CodeChunk}
		\begin{CodeInput}
R> get_probabilities_from_gips(g)
		\end{CodeInput}
	\end{CodeChunk}
	where \code{g} is the optimized \code{`gips`} object.  If one is interested only in the maximum a posteriori estimate, it is better to set \code{return\_probabilities = FALSE} in the \code{find_MAP()} function.
	
	Finally, to obtain the MLE of the covariance matrix in the invariant model found by \code{find_MAP}, one projects the empirical covariance matrix on a colored space corresponding to chosen permutation: 
	\begin{CodeChunk}
		\begin{CodeInput}
R> project_matrix(S, perm)
		\end{CodeInput}
	\end{CodeChunk}
	The first argument, \code{S}, is the $p\times p$ covariance matrix used in the \code{gips()} function. The \code{perm} is either a permutation or the 'gips' object that describes the cyclic permutation symmetry.

	\subsection{Real life example} \label{sec:brease_cancer}
	
	We obtained the results in this section with AMD EPYC 7413 on a single core, which took under 2 hours to compute.
	
	Let us present the capabilities of \pkg{gips} package using breast cancer data from \citetalias{Miller}.  
	Following the approach of \cite{HL08}, we consider a set of $p = 150$ genes and $n = 58$ samples with a mutation in the p53 sequence. We numbered the variables alphabetically. Since $p > n$, only parsimonious models can be fitted at all. Data is available in \code{GEOQuery} package from BioConductor. Code for downloading and minimal preprocessing is available in the ``Replication code''.
	We stress that the model space to search is here very large. It can be roughly estimated to be of magnitude $10^{250}$.
	
	\begin{CodeChunk}
		\begin{CodeInput}
R> Z <- breast_cancer
R> dim(Z)
		\end{CodeInput}
		\begin{CodeOutput}
[1]  58 150
		\end{CodeOutput}
	\end{CodeChunk}
	
	We note, that we have fewer observations than variables. Let us search for permutation symmetries. Create \code{`gips`} object and run \code{find_MAP()} function on it. We set \code{D_matrix = diag(p)} as in \cite[Section 4.1]{GIKM_SM}.
	\begin{CodeChunk}
		\begin{CodeInput}
R> S <- cov(Z)
R> g <- gips(S, 58, D_matrix = diag(p), was_mean_estimated = TRUE)
R> set.seed(2022)   
R> g_MAP <- find_MAP(g, max_iter = 150000, optimizer = "MH")
		\end{CodeInput}
	\end{CodeChunk}
	To acquire knowledge about the optimization process, we can call the \code{summary()} function on the object of the \code{`gips`} class.
	
	\begin{CodeChunk}
		\begin{CodeInput}
R> summary(g_MAP)
		\end{CodeInput}
		\begin{CodeOutput}
The optimized `gips` object.

Permutation:
 (1,10,83,61,69,37,137,106)(2,42,19,16,43,49,24,82,34,139,140,52,26,98,17,100,
97,145)(3,9,11,71,120,101,126,76)(4,8,89)(5,148)(6,30,149,107,65,78,60,127)
(7,133,36,95)(12,103,92,146,138,144,84,62,58,77,111,122,66,129,93,59,41,81,35,
64,86,117,63,150,70,75,108)(13,50,57,132,114,22,116,125,74,72,91,90,113,130,
124)(14,110,46,29)(15,51,56,48,53,25,45,119)(18,68,99)(20,79,21)(23,131,27,67,
38,128,147,112,102)(28,73,44,135,105,96,104,39)(31,40,118,115,143)(32,33,123,
134,121,88)(47,109,94,136)(141,142)

Log_posteriori:
 3626.114

Times more likely than starting permutation:
 7.865e+549

The number of observations:
 58

The mean in the `S` matrix was estimated.
Therefore, one degree of freedom was lost.
There are 57 degrees of freedom left.

n0:
 25

The number of observations is bigger than n0 for this permutation,
so the gips model based on the found permutation does exist.

The number of free parameters in the covariance matrix:
 611

BIC:
 8741.694

AIC:
 7482.764

-----------------------------------------------------------------------------
Optimization algorithm:
 Metropolis_Hastings

Number of log_posteriori calls:
 150000

Optimization time:
 1.428308 hours

Acceptance rate:
 0.00195333333333333

Log_posteriori calls after the found permutation:
 36814
		\end{CodeOutput}
	\end{CodeChunk}
	The resulting permutation consists of $C_\sigma = 24$ cycles and $n_0=25$. The dimension of the space $\mathcal{Z}_{\left<\sigma\right>}$, i.e., the number of free parameters in the covariance matrix, is $611$.
	
	We can interpret this result as an indication of hidden symmetry in genes and evidence that our procedure can be used as an exploratory tool for finding such
	symmetries. 
	
\begin{figure}[h]
		\centering
		\includegraphics[width=0.7\textwidth]{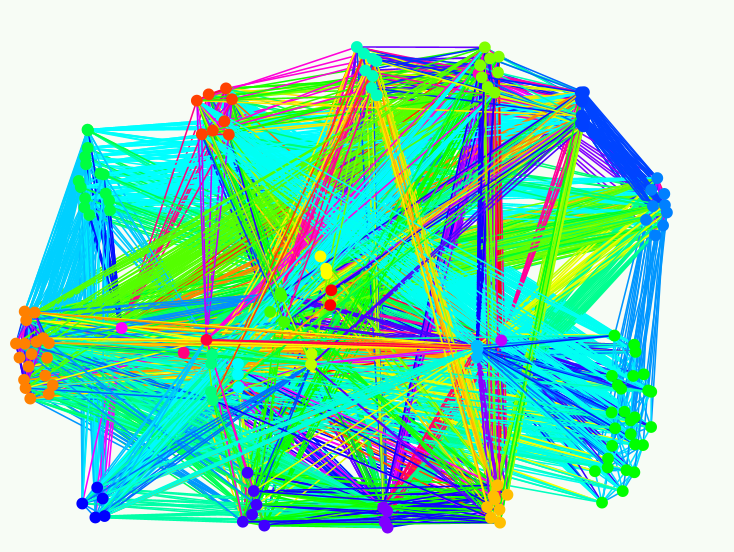}
		\caption{\small The undirected colored graph corresponds to constructed RCOP model. The variances of vertices with the same color are equal, and the covariances of edges with the same color are also equal. Additionally, entries of the precision matrix corresponding to missing edges are $0$.}
		\label{fig:breast_cancer}
	\end{figure}

	We also carry out the heuristic procedure introduced in \cite[Section 1.2]{GIKM} for finding a graphical model which is invariant under the above symmetry.  
	We threshold the entries of the partial
	correlation matrix at the level $\alpha = 0.05608621$ and construct undirected graph $G=(V,E)$ with $V=\{1,\ldots,p\}$ and $E=\{ \{i,j\}\colon i,j\in V, i\neq j\}$ defined by
	\[
	\{i,j\} \in E\quad\mbox{if and only if}\quad \frac{|k_{ij}|}{\sqrt{k_{ii}k_{jj}}} \geq \alpha,
	\]
	where $(k_{ij})$ are the entries of the estimated precision matrix $\hat K=\hat\Sigma^{-1}$. The constructed dependency graph is depicted in Figure~\ref{fig:breast_cancer}. The graph is non decomposable, it has $3324$ edges (compared to $p(p-1)/2 = 11175$ edges in the full graph with $p=150$ vertices) and the size of its biggest clique is $21$. We found the MLE of the covariance matrix in the corresponding colored graphical model using the \code{ggmfit()} function from the \pkg{gRim} package. Note that the maximum likelihood equation for this model can be solved by first taking appropriate averages of the elements in the Wishart matrix (projecting the empirical covariance matrix onto the corresponding colored space) and then solving the equations for corresponding graphical Gaussian model without symmetry restrictions \cite{HL08}.
	
	By deleting edges, the number of parameters was reduced from $611$ to $271$, resulting in a decrease in the log-likelihood from $-3130$ to $-3353$. This reduction leads to a lower BIC compared to the former model, decreasing from $8742$ to $7807$. These findings indicate that the simpler model provides a better description of the data and highlight the relevance of the entire approach.
	
	\subsection{Hyperparameters' influence}
	\label{sec:hyper_influence}
	
	We obtained the results in this section with AMD EPYC 7413 on 24 cores, which took under 5 minutes to compute.
	
	The Bayesian model introduced in Section \ref{sec:BM} depends on two parameters of the a priori distribution, a scalar $\delta$ and a matrix $D$. 
	In the following section, we present the effect of these hyperparameters on the a posteriori distribution. 
	Despite having an explicit formula for the quotient~(\ref{eq:defpp}), it is too complex to allow for direct analysis. Furthermore, this formula inherently depends on the data, which further complicates the study. 
	
	Therefore, drawing conclusions about the influence of hyperparameters on the method's outcome is difficult and must be done with caution. They directly influence the shape of the a posteriori distribution and, therefore, change both the theoretical MAP and the difficulty of the optimization problem.
	
	We consider only the low-dimensional setting $p=8$ because only then are we able to efficiently calculate posterior probabilities for all cyclic subgroups. On a standard PC, it takes about $4$ minutes to calculate the entire posterior distribution. Moreover, there is no rationale to suggest that the influence of the hyperparameters would be significantly different for larger $p$. Comparisons are conducted across three different scenarios.

	First, we generate an empirical covariance matrix \code{S} from the Wishart distribution on $\mathrm{Sym}(p;\mathbb{R})$ with the scale parameter $I_p$ and the shape parameter $p$. 
	Then, the true covariance matrices for three scenarios are defined as the projections (recall (\ref{pig})) of \code{S} onto the spaces invariant under the following cyclic permutation subgroups:
	\begin{description}
		\item[no structure:] $\left<\mathrm{id}\right>$,
		\item[moderate structure:]  $\left<(1,2,3,4)\right>$,
		\item[large structure:] $\left<(1,2,\ldots,8)\right>$.
	\end{description}
	
\begin{figure}[h]
		\centering
		\includegraphics[width=0.31\textwidth]{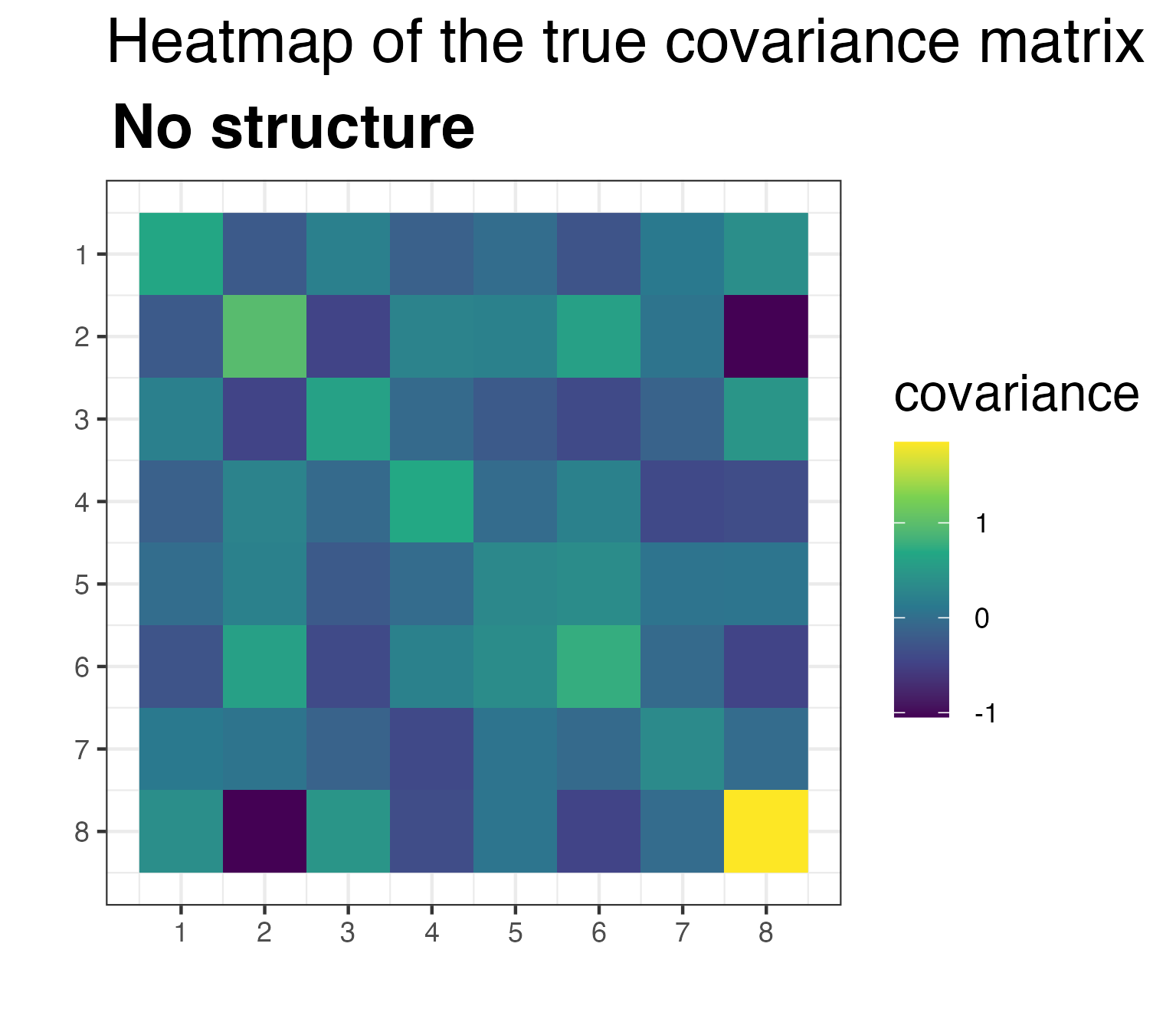}\quad
		\includegraphics[width=0.31\textwidth]{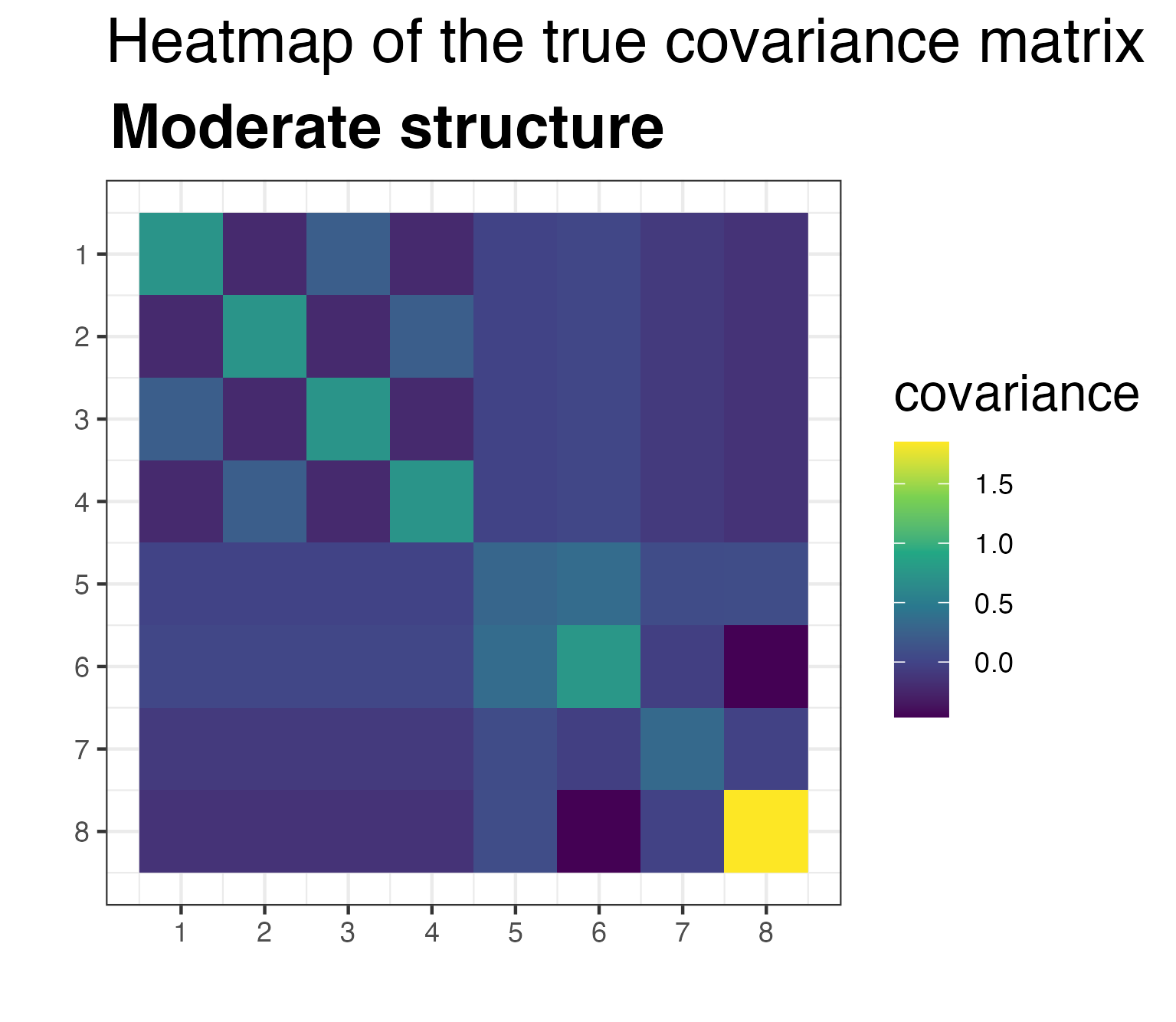}\quad
		\includegraphics[width=0.31\textwidth]{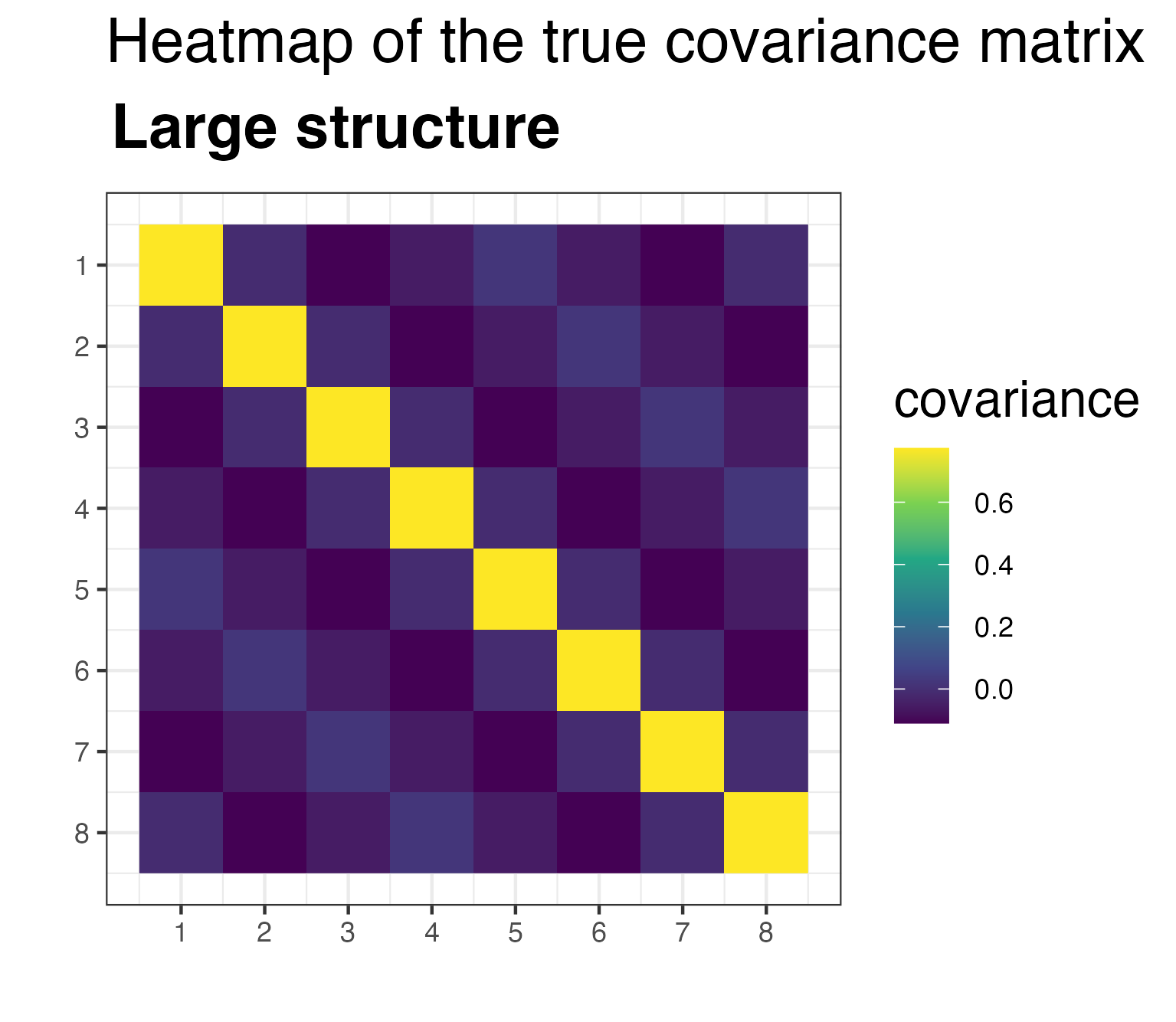}
		\caption{\small True covariance matrices corresponding to the three scenarios: left panel - no structure \code{S}, middle panel - moderate structure $\pi_{\left<(1,2,3,4)\right>}($\code{S}$)$, right panel - large structure $\pi_{\left<(1,2,\ldots,8)\right>}($\code{S}$)$.}
		\label{fig:covariance matrices}
	\end{figure}

	The number of free parameters (or just the dimension) of the model without structure is $p+p(p-1)/2=36$, while the model with moderate structure has dimension $17$, and the model with large structure has $5$ dimensions. For each of these three scenarios, we simulate $n=30$ samples from $\mathrm{N}_p(0,\Sigma)$, where $\Sigma$ is the true covariance matrix for a given scenario (see Figure~\ref{fig:covariance matrices}).

	In order to investigate the characteristics of the a posteriori distribution, we considered $\delta$ to be $3$ and $30$, and $D$ of the form $d \cdot I_p$ for $d\in\{0.1, 1, 10, 100\}$.  Recall that $\delta=3$ and $D=I_p$ are the default parameters. 
	Since there is no natural total ordering of cyclic subgroups, we identify each subgroup with the dimension of the model it generates, i.e., $\dim(\mathcal{Z}_\Gamma)$. In this way, for the sake of this analysis, all models that have the same dimension are merged.
	
\begin{figure}[h]
		\centering
		\includegraphics[width=0.45\textwidth]{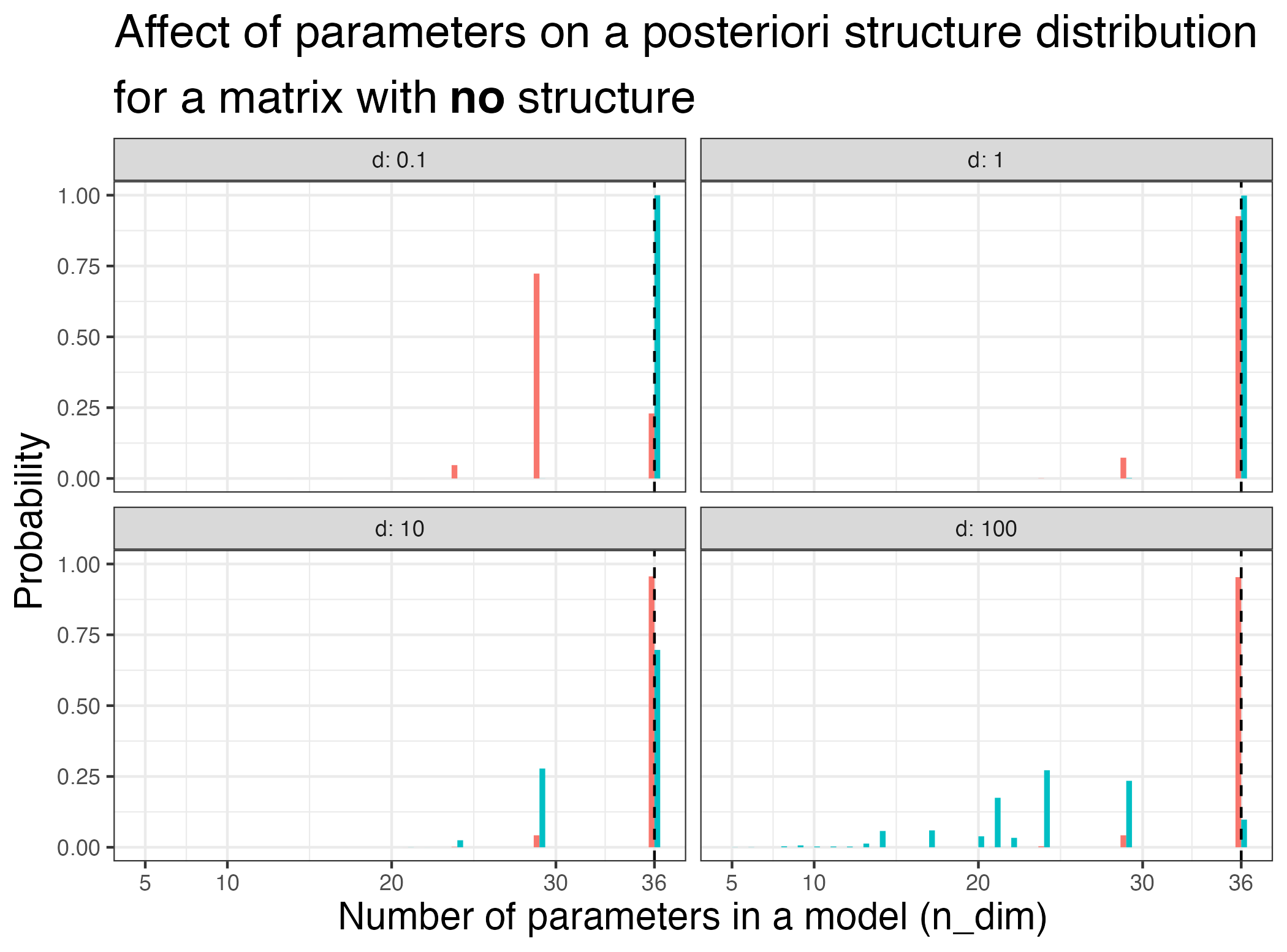}\quad
		\includegraphics[width=0.45\textwidth]{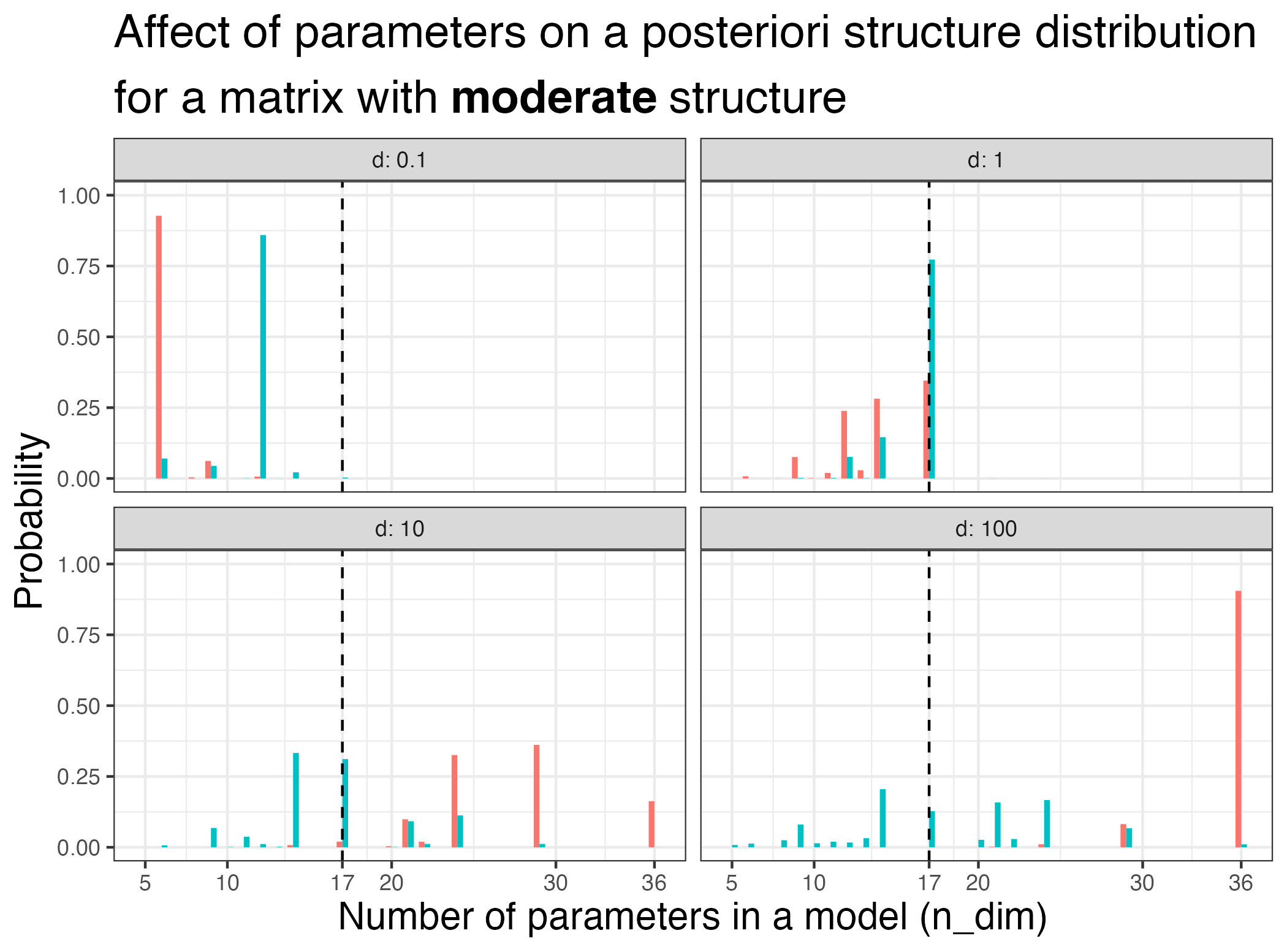}\\
		\includegraphics[width=0.50\textwidth]{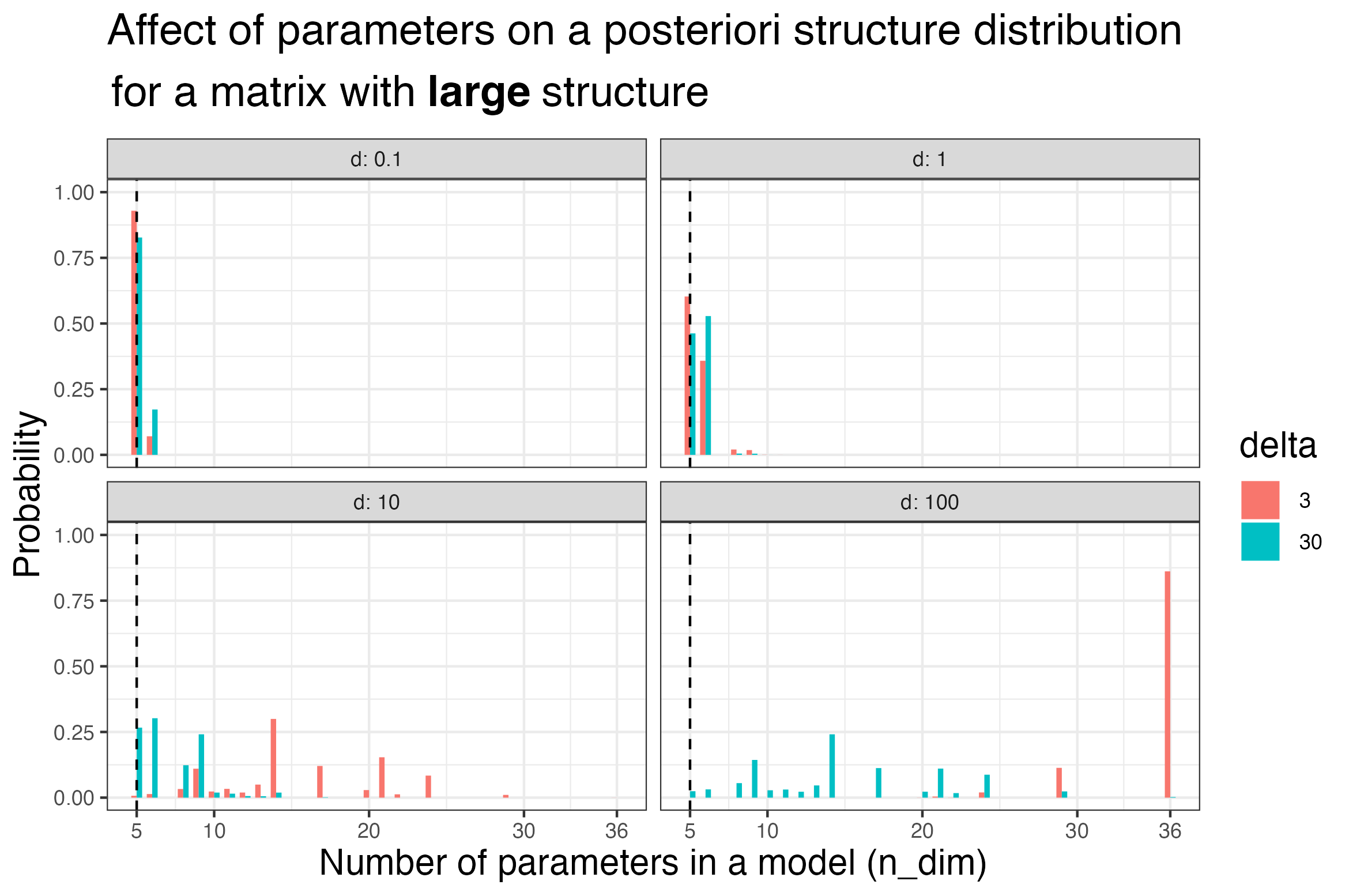}
		\caption{\small The posterior distribution (\ref{eq:exact}) for $\delta\in\{3,30\}$ and $D=d\cdot I_p$ for $d\in\{0.1, 1, 10, 100\}$ for three different scenarios. The black vertical line indicates the dimension of the true model.}
		\label{fig:posterior_distribution}
	\end{figure}

	The obtained distributions are shown at Figure~\ref{fig:posterior_distribution}.

	Comments:
	\begin{itemize}
		\item The influence of parameter $D=d\cdot I_p$ is best reflected in the moderate structure scenario. As the value of parameter $d$ decreases, larger symmetries (corresponding to lower-dimensional models) are preferred. 
		Increasing the parameter $\delta$ shifts the distribution and reduces the difference between probabilities and therefore increases the entropy of the distribution.
		\item On the other hand, the posterior distribution for large values of parameter $D=d\cdot I_p$ becomes independent of the data. In fact, the plots for $d=100$ are very similar across all the considered scenarios.
		\item In each of the scenarios, the Bayesian model correctly identifies the true model when $\delta=3$ and $D=1\cdot I_p$. The value of the default parameter $D$ for all scenarios is very similar, being equal to $0.77\cdot I_p$. This demonstrates the model's good properties for the default values of hyperparameters.
	\end{itemize}
	
	\subsection{Comparison with other methods}
	\label{sec:compwithother}
	
	We obtained the results in this section with AMD EPYC 7413 on 45 cores, which took around 1 hour 30 minutes to compute.
	
	As mentioned in the Introduction, 
	although there are no other software packages available for finding permutation symmetries in data, we have made the decision to compare the results of our model with canonical methods commonly used to tackle high-dimensional problems. In this section, we will compare method from \pkg{gips} package with methods implemented in \pkg{huge} (GLASSO) and \pkg{rags2ridges} (Ridge) packages. Both \pkg{huge} and \pkg{rags2ridges} are based on matrix penalization and include a hyperparameter $\lambda$, which controls the strength of the penalty. They also both have implemented hyperparameter search techniques, which we will utilize.
	
	\subsubsection{Methodology}
	
	We conducted the comparison across different sample sizes and across strengths of the symmetry structure of true covariance matrices, similarly as in the previous section. For $p=50$, we utilized matrices that are invariant under the following permutation subgroups: 
	\begin{description}
		\item[no structure:] $\left<\mathrm{id}\right>$,
		\item[moderate structure:]  $\left<(1,2,\ldots,25)\right>$,
		\item[large structure:] $\left<(1,2,\ldots,50)\right>$.
	\end{description}
	For each of these scenarios, we constructed the true covariance matrices in the following way: first, we sampled a positive definite matrix from the Wishart distribution on $\mathrm{Sym}(p; \mathbb{R})$ with a scale parameter of $I_p$ and a shape parameter of $p$. We then projected this matrix onto the colored space corresponding to a given scenario. Next, we thresholded the inverse of this matrix by setting $25\%$ of the off-diagonal entries with the smallest absolute values to zero.
	The inverse of such matrix served as our true covariance matrix. It is important to note that this approach does not always produce a positive definite matrix, which was indeed the case in the 'no structure' scenario. In this particular case, we added \code{0.1 * diag(p)} to the realization of the Wishart distribution. This adjustment ensured the construction of a proper covariance matrix.

	In the initial analysis, we also considered covariance matrices whose inverses did not contain any zeros. However, to our surprise, the results for matrices with and without zeros in their inverses were very similar for all the methods. Therefore, we decided to focus only on covariance matrices corresponding to nontrivial conditional dependence structures, as they are expected to favor likelihood penalization methods more.
	
	For each of the scenarios, we considered three different sample sizes $n\in \{10,20,40\}$.
	Therefore, we have a total of $3\cdot3 = 9$ settings for this experiment. The comparison method for each setting is as follows:
	\begin{enumerate}
		\item[1)] Fix a sample size $n$ and true covariance matrix $\Sigma$.
		\item[2)] Generate a sample $Z$ from $\mathrm{N}_p(0, \Sigma)$ of size $n$.
		\item[3)] Estimate the covariance matrix using:
		\begin{itemize}
			\item \pkg{gips} package: \code{find_MAP()} function with \code{optimizer = "MH"} and \code{n_iter = 300000},
			\item \pkg{rags2ridges} package: \code{ridgeP()} function with \code{lambda} parameter found by the \code{optPenalty.kCVauto()} function with \code{lambdaMin = 0.001}, \code{lambdaMax = 100} range,
			\item \pkg{huge}: function \code{huge()} with parameters \code{method = "glasso"} and \code{nlambda = 40} and \code{llambda.min.ratio = 0.02} and the tuning parameter \code{lambda} was selected using default parameters of \code{huge.select()} function (rotation information criterion).
		\end{itemize}
		\item[4)] Record the log-likelihood and evaluate estimation using the Frobenius norm.
		\item[5)] Repeat 2)-4) 10 times and aggregate results.
	\end{enumerate}
	
	Recall that the Frobenius norm of a $p\times p$ matrix $M=(m_{ij})_{i,j}$ is defined by 
	\[
	\|M\|_F = \sqrt{\sum_{i=1}^p\sum_{j=1}^p |m_{ij}|^2}.
	\]
	When $M$ is a difference between the true covariance matrix and its estimate, $\|M\|_F^2$ is proportional to the mean squared error (MSE).
	
	\subsubsection{Results}
	
\begin{figure}[h]
		\centering
		\includegraphics[width=0.9\textwidth]{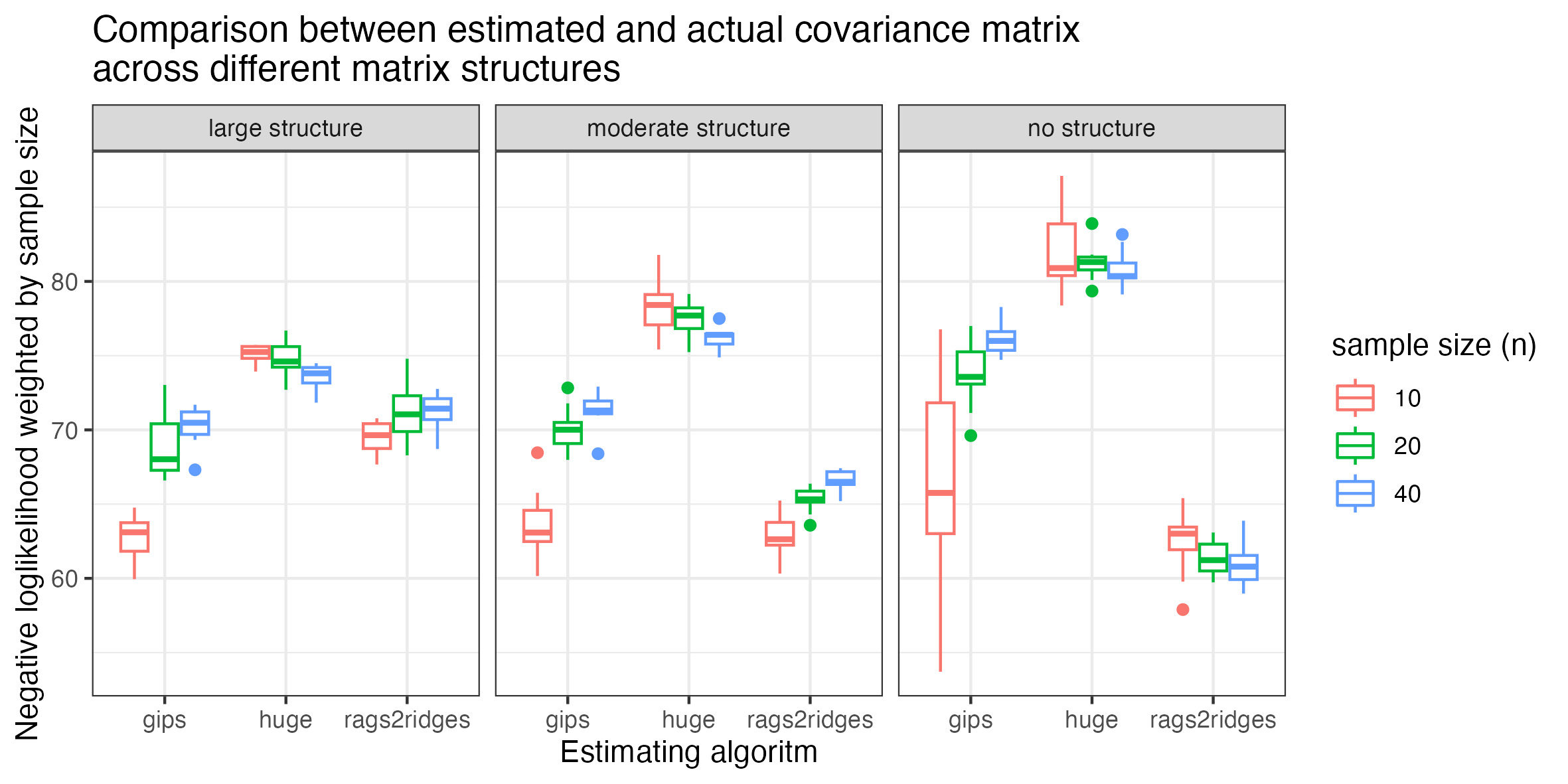}
		\caption{\small Negative log-likelihoods of covariance matrix estimations. $10$ runs for each configuration.}
		\label{fig:loglik}
	\end{figure}

	Each of the three methods produces an estimator of the covariance matrix. We calculate the corresponding (negative) log-likelihood in our Gaussian model and present the results in Figure~\ref{fig:loglik}.
	
	Comments:
	\begin{itemize}
		\item In all configurations, the \pkg{gips} method outperformed \pkg{huge}. However, it should be emphasized that the main purpose of the GLASSO method is the model selection within graphical models rather than the estimation of the covariance matrix. Typically, when possible, estimation is performed within the selected model and such an approach leads to systematically smaller bias.
		\item The comparison with \pkg{rags2ridges} is more interesting, as the \pkg{gips} method yielded weaker results when there was no structure and better results when the symmetry structure was large. This behavior was expected as \pkg{gips} is designed to look for these structures in the data.
		\item We can see that the results of \pkg{gips} were very unstable when there was no structure in the underlying ground truth matrix. This behaviour is expected as \pkg{gips} will more likely find some non-existing structure when $n$ is much smaller than $p$.  Each method gains stability when the sample size increases.
	\end{itemize}
	
	
	
\begin{figure}[h]
		\centering
		\includegraphics[width=0.9\textwidth]{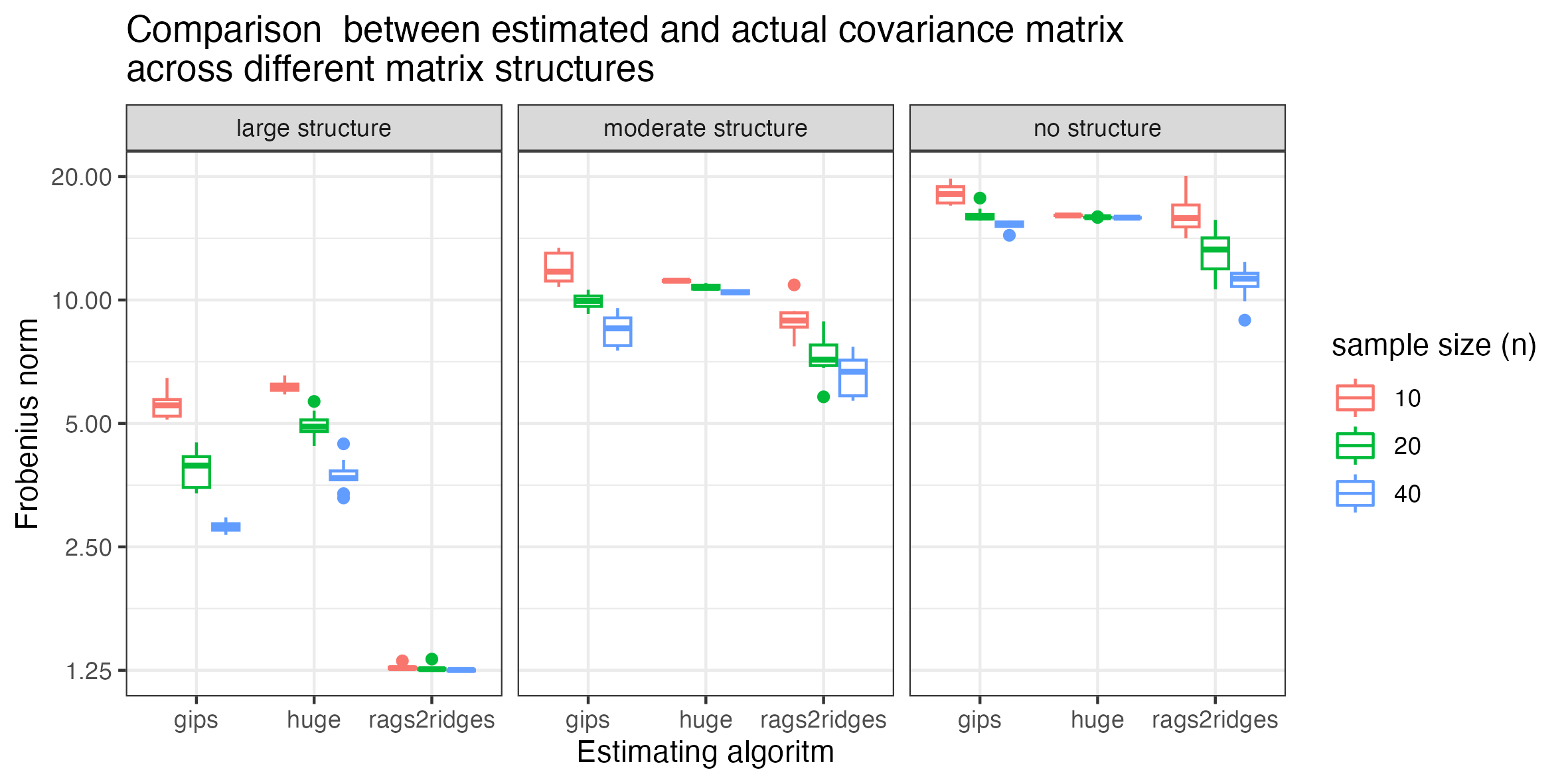}
		\caption{\small Frobenius norm (on a logarithmic scale) of the difference of the estimate and the true covariance matrix. $10$ runs for each configuration.}
		\label{fig:frob}
	\end{figure}

	The Frobenius norm of the difference of the estimate and the true covariance matrix is shown in Figure~\ref{fig:frob} with logarithmic Y axis.
	
	\newpage
	Comments:
	\begin{itemize}
		\item When there is no symmetry structure in the data, the three methods considered generate estimates that are very similar in terms of the Frobenius norm. 
		\item Generally, the bigger the symmetry structure, the better the quality of the estimation. Both \pkg{gips} and \pkg{huge} perform similarly in all scenarios, while \pkg{rags2ridges} performs significantly better in the scenario with a large structure.
	\end{itemize}
	
	We acknowledge that the proposed method of comparison is not systematic enough to draw conclusions in full generality. 
	On one hand, the results act in support of the theory: \pkg{gips} method is a viable choice if we suspect, that the true matrix has some structure. On the other hand, it is difficult to recognise it post-hoc by comparing the method's performance using log-likelihood (or possibly other measures) usable in real-world cases, when the true covariance matrix is unknown.
	
	From a practical perspective, it is worth repeating, that the \pkg{gips} method's output provides not only a projected covariance matrix but also an interpretation in the language of permutation symmetries of the data. 
	
	Finally, we note that both \pkg{huge} and \pkg{rags2ridges} methods execute within a few seconds, while it takes approximately $20$ minutes to run the \pkg{gips} method for one scenario for $p=50$ and $300\,000$ iterations of the Metropolis-Hastings algorithm.
	
	\section{Summary and discussion} \label{sec:summary}
	
	In this paper, we have presented \pkg{gips}, an \proglang{R} package for learning permutation symmetry (i.e., identifying exchangeable variables) from Gaussian multivariate sample. Its usage revolves around setting up the main object of class \code{`gips`} and then calling main functions \code{find_MAP} and \code{project_matrix}. Implementations of methods well known to \proglang{R} users, such as \code{summary} or \code{plot}, are also available. The applicability of \pkg{gips} spans a diverse array of scenarios where permutation symmetries are anticipated, as detailed in Section \ref{sec:motivation}. It also proves advantageous in a high-dimensional setting, where conventional methods fail.
	
	Our model provides interpretable results which are competitive to popular dimensionality reduction and covariance matrix estimation methods in Gaussian models. It excels when the underlying probability distribution has a rich symmetry structure, since it is able to detect and use that characteristic directly.
	We emphasize that, in the field of model selection methods for colored graphs, there is currently no competition for our package. Known methods are, to the best of our knowledge, not implemented in a publicly available package and the model presented in \cite{GIKM} and implemented in \pkg{gips} is the only one that allows for the search of permutation symmetries.
	
	The \pkg{gips} package is under active maintenance and will continue to be developed to incorporate more advanced features. One potential avenue for future development is the inclusion of a model selection procedure within Gaussian graphical models that are invariant under permutation symmetry (RCOP models). By providing this package, we intend to stimulate the discussion about the concept of permutation symmetries and the investigation of its potential applications for statisticians and the \proglang{R} community. We invite everyone to a discussion about potential directions of development at \url{https://github.com/PrzeChoj/gips/issues}.

	
	\section*{Computational details}

	The ``Replication code'' is available at \url{https://github.com/PrzeChoj/gips_replication_code}.
	
	The results in this paper were obtained using
	\proglang{R}~4.2.1 with the	\pkg{gips}~1.2.1 package. \proglang{R} itself
	and most packages used are available from the Comprehensive \proglang{R} Archive Network (CRAN) at \url{https://CRAN.R-project.org/}. The only exception is package \pkg{GEOquery} 2.66.0 \cite{GEOquery} that was used to obtain the data for Section \ref{sec:brease_cancer}. This package is available on Bioconductor repository \citetalias{bioconductor} at \url{http://www.Bioconductor.org/}.
	For \pkg{gips}'s dependencies, we used \pkg{numbers} 0.8-5 \cite{numbers}, \pkg{permutations} 1.1-2 \cite{permutations}, \pkg{rlang} 1.1.1 \cite{rlang}, \pkg{utils} 4.2.2 \cite{utils}.
	
	For packages in Section \ref{sec:compwithother}, we used \pkg{rags2ridges} 2.2.6, \pkg{huge} 1.3.5.
	
	The remaining packages used are \pkg{Biobase} 2.58.0, \pkg{GEOquery} 2.66.0, \pkg{BiocManager} 1.30.21, \pkg{MASS} 7.3-60 \cite{MASS}, \pkg{ggplot2} 3.4.2 \cite{ggplot2}, \pkg{magrittr} 2.0.3 \cite{magrittr}, \pkg{parallel} 4.2.2, \pkg{dplyr} 1.1.2 \cite{dplyr}, \pkg{stringi} 1.7.12 \cite{stringi}, \pkg{gRim} 0.2.10 \cite{gRim}.
	
	For producing the Figure~\ref{fig:breast_cancer} we used Cytoscape v3.10.0 \cite{shannon2003cytoscape}.
	
	\section*{Acknowledgments}
	Research was funded by (POB Cybersecurity and Data Science) of Warsaw University of
	Technology within the Excellence Initiative: Research University.
	
	This research was carried out with the support of the Laboratory of Bioinformatics and Computational Genomics and the High Performance Computing Center of the Faculty of Mathematics and Information Science Warsaw University of Technology.

	\bibliography{biblio}

	\newpage
	
	\begin{appendix}
		
		\section{Example to Section 1.4}
		\label{app:general}
		
		The standard PC can run all the code in this appendix within 2 seconds.
		
		Consider an i.i.d.~sample $(\texttt{Z}^{(i)})_{i=1}^{\texttt{n}}$ from $\mathrm{N}_{\texttt{p}}(0, I_{\texttt{p}})$ for $\texttt{p}=4$ and $\texttt{n}=50$ and let \code{S} be its empirical covariance matrix. The distribution of \code{Z} is clearly invariant under any permutation. Let us examine the output of \pkg{gips}.
		\begin{CodeChunk}
			\begin{CodeInput}
R> p <- 4; n <- 50
R> set.seed(2022); Z <- matrix(rnorm(n * p), ncol = p);  S <- cov(Z)
R> g <- gips(S, n)
R> g_MAP <- find_MAP(g, optimizer = "BF",
+    return_probabilities = TRUE, save_all_perms = TRUE
+  )
R> get_probabilities_from_gips(g_MAP)
			\end{CodeInput}
			\begin{CodeOutput}
   (1,3,2,4)    (1,2,3,4)    (1,2,4,3)      (1,3,4)      (2,3,4)      (1,2,4) 
2.542477e-01 2.393022e-01 2.124555e-01 1.837534e-01 6.234428e-02 2.517374e-02 
     (1,2,3)   (1,4)(2,3)   (1,3)(2,4)   (1,2)(3,4)        (3,4)        (1,4) 
1.971380e-02 9.287411e-04 6.006721e-04 4.634810e-04 3.743338e-04 2.542462e-04 
       (1,3)        (2,4)        (2,3)        (1,2)           () 
1.811636e-04 8.977424e-05 8.207691e-05 3.465056e-05 2.263418e-07 
			\end{CodeOutput}
		\end{CodeChunk}
		We observe that the symmetries with the highest probability correspond to the long cycles, and these probabilities are very close to each other. This suggests that the data is invariant under each of these symmetries. The only model invariant under these three symmetries is the full-symmetry model, which is invariant under any permutation (both the diagonal and off-diagonal of the covariance matrix are constant).

		\section{Formulas for structure constants} \label{app:technical}
		In this appendix, we outline the steps required to find the ingredients necessary for the calculation of the normalizing constants 
		\[
		I_{\Gamma} (\delta,D)=  \int_{\mathcal{P}_\Gamma} \mathrm{Det}(k)^{(\delta-2)/2} e^{- \tfrac12 \mathrm{Tr}[D\cdot k]} dk,\qquad \delta>1,\,D\in \mathrm{Sym}^+(p;\mathbb{R}). 
		\]
		for arbitrary cyclic subgroup $\Gamma$. These constants are indispensable for our model selection procedure.  
		
		We note that the formulas for normalizing constants for an arbitrary subgroup $\Gamma\subset\mathfrak{S}_p$ are presented in \cite{GIKM}. Here, we specialize these formulas to cyclic subgroups, which allows for significant simplification.
		
		Let $p_i$ be the length of the $i$-th cycle in the cyclic decomposition of $\sigma\in\mathfrak{S}_p$, and let $\{i_1, \dots, i_{C_\sigma}\}$ be a complete system of representatives of the cycles of $\sigma$. Furthermore, let $(e_i)_{i=1}^p$ be the standard basis of $\mathbb{R}^p$.
		\begin{enumerate}
			\item For $c=1,\ldots,C_\sigma$, calculate $v^{(c)}_1, \dots, v^{(c)}_{p_c} \in \mathbb{R}^p$ as 
			\begin{align*}
				v^{(c)}_1 &:= \sqrt{\frac{1}{p_c}} \sum_{k=0}^{p_c-1} e_{\sigma^k(i_c)}, \\
				v^{(c)}_{2\beta} 
				&:= \sqrt{\frac{2}{p_c}} \sum_{k=0}^{p_c-1} \cos \Bigl( \frac{2\pi \beta k}{p_c} \Bigr) e_{\sigma^k(i_c)} \qquad (1 \le \beta < p_c/2),\\
				v^{(c)}_{2\beta+1} 
				&:= \sqrt{\frac{2}{p_c}} \sum_{k=0}^{p_c-1} \sin \Bigl( \frac{2\pi \beta k}{p_c} \Bigr) e_{\sigma^k(i_c)}
				\qquad (1 \le \beta < p_c/2),\\
				v^{(c)}_{p_c} &:= \sqrt{\frac{1}{p_c}} \sum_{k=0}^{p_c-1} \cos (\pi k) e_{\sigma^k(i_c)}
				\qquad\quad\,\,\, (\mbox{if }p_c \mbox{ is even}).
			\end{align*}
			\item Construct an orthogonal matrix $U_\Gamma$ by arranging column vectors $\{v^{(c)}_k\}$, $1 \le c \le C_\sigma$, $1\le k \le p_c$, in the following way:
			we put $v^{(c)}_k$ earlier than $v^{(c')}_{k'}$
			if\\ 
			{\rm (i)} $\frac{[k/2]}{p_c} < \frac{[k'/2]}{p_{c'}}$, or\\
			{\rm (ii)} $\frac{[k/2]}{p_c} = \frac{[k'/2]}{p_{c'}}$ 
			and $c<c'$, or\\
			{\rm (iii)} $\frac{[k/2]}{p_c} = \frac{[k'/2]}{p_{c'}}$ and $c = c'$ and $k$ is even and $k'$ is odd.
			\item Let $N$ be the order of $\Gamma$. For $\alpha=0,1,\ldots,\lfloor\frac{N}{2}\rfloor$ calculate
			\begin{align*}	
				r_\alpha^\ast &= \#\{c\in\{1,\ldots,C_\sigma\}\colon\alpha\, p_c  \mbox{ is a multiple of }N\},\\
				d_\alpha^\ast &= \begin{cases} 1 & (\alpha = 0 \mbox{ or }N/2), \\ 2 & \mbox{(otherwise)}. \end{cases}
			\end{align*}
			n the definition of $r_\alpha^\ast$, we treat $0$ as a multiple of $N$, and thus $r_0^\ast=C_\sigma$. 
			
			Then, we set $L=\#\{\alpha\colon r_\alpha^\ast>0\}$,
			$r = (r_\alpha^\ast\colon r_\alpha^\ast>0)$ 
			and
			$d = (d_\alpha^\ast\colon  r_\alpha^\ast>0)$.
			The parameters $(r_i,d_i)_{i=1}^L$ are called the structure constants.
		\end{enumerate}
		
		The constructed orthogonal matrix $U_\Gamma$ possesses a notable property. According to \cite[Theorem 5]{GIKM}, it performs a block decomposition of the colored space $\mathcal{Z}_\Gamma$ in the following sense: for each $S\in\mathcal{Z}_\Gamma$, we have
		\begin{equation}
			\label{genform}
			U_\Gamma^\top\cdot S\cdot U_\Gamma
			= 
			\begin{pmatrix} 
				x_1 & & \\
				& \ddots & \\
				& & x_L
			\end{pmatrix},
		\end{equation}
		where $x_i\in\mathrm{Sym}(r_i\,d_i;\mathbb{R})$, $i=1,\ldots,L$. 
		
		For any $S\in\mathrm{Sym}^+(p; \mathbb{R})$ and $\delta\in\mathbb{R}$, we define a function 
		\begin{align*}
			\gamma_\Gamma(S,\delta) = \prod_{i=1}^L \mathrm{Det}(x_i)^{-(\delta+r_i-3)/2-1/d_i},
		\end{align*}
		where $x_i\in\mathrm{Sym}(r_i\,d_i;\mathbb{R})$ are the diagonal blocks of a decomposition (\ref{genform}) of $\pi_\Gamma(S)$ (recall (\ref{pig})).
		
		Finally, by \cite[Theorem 9]{GIKM}, integral $I_\Gamma(\delta,D)$ is convergent if
		$(\delta-2)/2>\max_{i=1}^L \{-1/d_i\}$ and $D$ is positive definite. The expression $\max_{i=1}^L \{-1/d_i\}$ equals $-1/2$ unless $L=1$ (which corresponds to the trivial subgroup $\{\mathrm{id}\}$), in which case it is equal to $-1$. Thus, for all $\delta>1$ and $D\in\mathcal{P}_\Gamma$ we have
		\begin{align*}
			I_\Gamma(\delta,D)= e^{- A_\Gamma (\delta-2)/2 - B_\Gamma} \gamma_\Gamma\left(\frac12D,\delta\right)
			\prod_{i=1}^L \Gamma_{i}\left(1+ d_i (\delta+r_i-3)/2\right),
		\end{align*}
		where 
		\begin{gather*}
			A_\Gamma = \sum_{i=1}^L r_i\, d_i\log d_i, \qquad 
			B_\Gamma = \frac12\sum_{i=1}^L r_i (1+(r_i-1)d_i/2)\log d_i,\\
			\Gamma_{i}(\lambda)= 
			(2\pi)^{r_i(r_i-1) d_i/4}\prod_{k=1}^{r_i}\Gamma(\lambda-(k-1)d_i/2).
		\end{gather*} 
		
	\end{appendix}

\end{document}